\documentclass[12pt,journal,draft,a4paper,onecolumn]{IEEEtran}
\usepackage[cmex10]{amsmath}
\def\Mult{\mathop{\rm Mult}\nolimits}
\def\Irr{\mathop{\rm Irreducible}\nolimits}
\newtheorem{Lemma}{Lemma}
\newtheorem{Theorem}{Theorem}
\newtheorem{Corollary}{Corollary}
\begin{document}
\title{A novel method for computation of the discrete Fourier transform over 
characteristic two finite field of even extension degree}
\author{Sergei~V.~Fedorenko
}
\maketitle

\begin{abstract}
A novel method for computation of the discrete Fourier transform 
over a finite field with reduced multiplicative complexity is described.
If the number of multiplications is to be minimized, 
then the novel method for the finite field of even extension degree
is the best known method of the discrete Fourier transform computation.
A constructive method of constructing for a cyclic convolution over a finite field is introduced.
\end{abstract}


\section{Introduction}

The Reed--Solomon codes are used to correct errors 
in digital storage and communication systems, 
and for many other applications.
The discrete Fourier transform (DFT) over a finite field 
can be applied for encoding and decoding of the Reed--Solomon codes.

Let us consider $a + b$ (or $a \times b$) to be an addition (a multiplication) 
only if both summands (factors) lie in the original field \cite{Blahut85}; 
that is, operations in the prime field are neglected \cite{Blahut83}. 
The main problem in this paper is to reduce the multiplicative
(primarily) and additive (secondarily)
complexity of the DFT computation over a finite field.
In previous author's publications 
\cite{Fedorenko02, Fedorenko03, Fedorenko06, Fedorenko08}
the new methods of the DFT computation 
(the cyclotomic and recurrent algorithms) are introduced.
There are several papers with reduced additive complexity 
of the cyclotomic DFT (for example, \cite{Chen,Wu,Bellini}).

In the author's recent paper \cite{Fedorenko11} an idea and an example for 
a novel method with reduced multiplicative complexity 
of the DFT computation is first proposed.
This novel method is based on replacing a cyclic convolution 
by a multipoint evaluation.
For construction of a multipoint evaluation we use 
the novel modification of the Goertzel--Blahut algorithm
such that the polynomials on divisions levels 
have a small number of the coefficients in $GF(2^m) \backslash GF(2)$ 
(almost all coefficients are 0 or 1).
The algorithm of the DFT computation is valid for all composite extension degree $m$.
Without loss of generality, we can assume that $m$ is even.

The paper is organized as follows. 
In section \hbox{II}, we give basic notions and definitions.
In section \hbox{III}, we describe well-known algorithms for the DFT computation,
the novel modification of the Goertzel--Blahut algorithm, 
and theorem about the multiplicative complexity coincidence of well-known algorithms.
In section \hbox{IV}, we propose and prove 
the novel algorithm for the computation of a multipoint evaluation.
In section \hbox{V}, we present the novel method for the DFT computation,
which is based on replacing a cyclic convolution by a multipoint evaluation.
Section \hbox{VI} presents examples illustrating the developed techniques.

\section{Basic notions and definitions}
  
The discrete Fourier transform of length $n$ 
of a vector $f = (f_i)$, \, $i \in [0,n-1]$, ${n \mid (2^m-1)}$, 
in the field $GF(2^m)$
is the vector $F = (F_j)$, 
$$F_j = \sum_{i=0}^{n-1} f_i \alpha^{ij}, \quad j \in [0,n-1],$$ 
where $\alpha$ is an element of order $n$ in $GF(2^m)$ (a transform kernel).
Let us write the DFT in matrix form: 
$$F=Wf,$$
where $W = (\alpha^{ij})$, \, $i,j \in [0,n-1]$, 
is a Vandermonde matrix.

We further assume that the length of the $n$-point Fourier transform over 
$GF(2^m)$ is $n=2^m-1$.

Every vector $f$ is associated with a polynomial 
$f(x) = \sum_{i=0}^{n-1} f_i x^i$,
and we have $F_j=f(\alpha^j)$.

The field of the computation is the finite field $GF(2^m)$.

Let $\alpha$ be a primitive element of the field $GF(2^{m})$.

Definition. The binary conjugacy class of $GF(2^m)$ is: 

$$\left(\alpha^{c_k},\alpha^{c_k2},\alpha^{c_k2^2}, \ldots,\alpha^{c_k2^{m_k-1}}\right),$$
where $\alpha$ is a primitive element of the field $GF(2^{m})$, 
$\alpha^{c_k}$ is a generator of the $k$th binary conjugacy class, 
$m_k$ is a cardinality of the $k$th binary conjugacy class, $m_k \mid m$.
More generally the $q$-ary conjugacy class of $GF(2^m)$ is:
$$\left(\alpha^{c_k},\alpha^{c_kq},\alpha^{c_kq^2}, \ldots,\alpha^{c_kq^{m_k-1}}\right).$$

Let $l$ be the number of binary conjugacy classes of $GF(2^m)$.

Definition.  The minimal polynomial over $GF(q) \subset GF(2^m)$ of $\beta \in GF(2^m)$ is the lowest degree monic polynomial $M(x)$ with coefficients from $GF(q)$ such that $M(\beta) = 0$. 

Let $M_k(x)$ be the minimal polynomial over $GF(2)$  of the element $\alpha^{c_k}$.

Let $M_{k,i}(x)$ be the minimal polynomial over $GF(2^{m/2})$ of the element $\alpha^{c_k 2^i}$.

The matrix

$$
\begin{pmatrix}
 \alpha_1^0           & \alpha_1^1           & \alpha_1^2            & \cdots & \alpha_1^{t-1} \\ 
 \alpha_2^0           & \alpha_2^1           & \alpha_2^2            & \cdots & \alpha_2^{t-1} \\
 \alpha_3^0           & \alpha_3^1           & \alpha_3^2            & \cdots & \alpha_3^{t-1} \\
  \cdots                  & \cdots                    &  \cdots                    & \cdots &  \cdots               \\
 \alpha_t^0           & \alpha_t^1             & \alpha_t^2             & \cdots & \alpha_t^{t-1} \\
\end{pmatrix}
= \left(\alpha_i^{j-1}\right), \quad i,j \in [1,t],
$$
is called a Vandermonde matrix.

The matrix

$$
\begin{pmatrix}
 \alpha_1^1           & \alpha_1^q           & \alpha_1^{q^2}            & \cdots & \alpha_1^{q^{t-1}} \\ 
 \alpha_2^1           & \alpha_2^q           & \alpha_2^{q^2}            & \cdots & \alpha_2^{q^{t-1}} \\
 \alpha_3^1           & \alpha_3^q           & \alpha_3^{q^2}            & \cdots & \alpha_3^{q^{t-1}} \\
  \cdots                  & \cdots                    &  \cdots                            & \cdots  &  \cdots                       \\
 \alpha_t^1           & \alpha_t^q            & \alpha_t^{q^2}              & \cdots & \alpha_t^{q^{t-1}} \\
\end{pmatrix}
= \left(\alpha_i^{q^{j-1}}\right), \quad i,j \in [1,t],
$$
is called a Moore matrix \cite{Moore}.

The transpose of the Vandermonde (Moore) matrix we called 
the Vandermonde (Moore) matrix, too.

If the matrix is a Vandermonde matrix and a Moore matrix at the same time, 
then it is called a Moore--Vandermonde matrix.

A basis $\left(\beta^0, \beta^1, \beta^2, \ldots, \beta^{m_k-1}\right)$ 
of $GF(2^{m_k})$ over $GF(2)$ is a polynomial basis for the subfield $GF(2^{m_k}) \subset GF(2^m)$.

Let
$$\left(\gamma_k^{2^0}, \gamma_k^{2^1}, \gamma_k^{2^2}, \ldots, \gamma_k^{2^{m_k-1}}\right)$$
be a normal basis for the subfield 
$GF(2^{m_k}) \subset GF(2^m)$.

A circulant matrix, or a circulant, 
is a matrix, each row of which is obtained from the preceding
row by a left (right) cyclic shift by one position. 
Let us denote by $L_k$ an $m_k \times m_k$ 
circulant the first row of which is a normal basis. 
We call it a basis circulant \cite{Fedorenko06}:
$$
L_k = \begin{pmatrix}
\gamma_k^{2^0}      &\gamma_k^{2^1} &\ldots&\gamma_k^{2^{m_k-1}} \\
\gamma_k^{2^1}      &\gamma_k^{2^2} &\ldots&\gamma_k^{2^0}       \\
\ldots            &\ldots       &\ldots&\ldots             \\
\gamma_k^{2^{m_k-1}}&\gamma_k^{2^0} &\dots &\gamma_k^{2^{m_k-2}}
\end{pmatrix}
= \left(\gamma_k^{2^{i+j}}\right), \quad i,j \in [0,m_k-1].
$$

The basis circulant matrix is a Moore matrix for $q=2$.

Definition. A linearized polynomial over $GF(2^m)$ is a polynomial of the form
$$ L(x)=\sum_i a_i x^{2^i}, \quad a_i\in GF(2^m). $$
It can be easily seen that $L(a+b)=L(a)+L(b)$ holds for linearized polynomials.

\section{The algorithms for the DFT computation}
\subsection{The Goertzel--Blahut algorithm}

We consider the Blahut modification for the DFT computation
over finite fields \cite{Blahut83} of Goertzel's algorithm \cite{Goertzel}.

The first step of the Goertzel--Blahut algorithm is a long division of 
$f(x)$ by each minimal polynomial $M_k(x)$:
$$
\begin{cases}
f(x) = M_k(x) q_k(x) + r_k(x),   \\
\deg r_k(x) < \deg M_k(x) = m_k, \\
k \in [0,l-1], 
\end{cases}
$$
where $r_k(x) = \sum_{j=0}^{m_k-1} r_{j,k} x^j$, 
$l$ is the number of binary conjugacy classes.

The second step of the Goertzel--Blahut algorithm is 
to evaluate the remainder at each element of the finite field:
$$
\begin{cases}
F_i = f(\alpha^i) = r_k(\alpha^i) = \sum_{j=0}^{m_k-1} r_{j,k} \alpha^{ij}, \\
i \in [0,n-1], 
\end{cases}
$$
where an element $\alpha^i$ is a root of the minimal polynomial $M_k(x)$.

The matrix form for the first step is
$$
\begin{pmatrix}
\begin{tabular}{ll}
$(r_{j,0})$,  & $j \in [0,m_0-1]$     \\ \hline
$(r_{j,1})$,  & $j \in [0,m_1-1]$     \\ \hline
              & $\cdots$              \\ \hline
$(r_{j,l-1})$,& $j \in [0,m_{l-1}-1]$ \\
\end{tabular}
\end{pmatrix}
=
Rf,
$$
where $R$ is an $n \times n$ binary matrix.

The matrix form for the second step is
$$
\begin{tabular}{l}
$
\begin{pmatrix}
(F_{c_k 2^i}), \, i \in [0,m_k-1]
\end{pmatrix}
=
V_k 
\begin{pmatrix}
(r_{i,k}), \, i \in [0,m_k-1]
\end{pmatrix},$ 
$k \in [0,l-1],$ 
\end{tabular}
$$
where $V_k = \left(\alpha^{c_k j 2^i}\right)$, \, $i,j \in [0,m_k-1]$,
is a Moore--Vandermonde matrix.

The Goertzel--Blahut algorithm can be written in matrix form 
\cite{Fedorenko09, Blahut10} as follows:
\begin{equation}
\pi F = V R f,   
\end{equation}
where $\pi$ is an $n \times n$ permutation matrix, 
\begin{equation}
V = 
\begin{pmatrix}
V_0 &     &        &          \\
    & V_1 &        &          \\
    &     & \ddots &          \\
    &     &        & V_{l-1}  \\
\end{pmatrix}
\end{equation}
is an $n \times n$ block diagonal matrix composed 
of Moore--Vandermonde matrices $V_k$,\, $k \in [0,l-1]$.

\subsection{The cyclotomic algorithm}

The cyclotomic algorithm for the DFT computation \cite{Fedorenko03} is based on representing 
the initial polynomial $f(x)$ as a
sum of linearized polynomials (cyclotomic decomposition of the polynomial), 
finding their values in a set of base points (cyclic convolution), 
and computing the resulting vector as a linear combination
of these values with coefficients from a prime field 
(multiplication of a binary matrix by a vector).

The cyclotomic algorithm consists of the following steps:

0) decomposing an original polynomial into a sum of linearized polynomials
$$ f(x)=\sum L_k(x);$$

1) evaluating linearized polynomials at a set of basis points
$$\{ L_k(\gamma_i) \};$$

2) components of the Fourier transform are computed as linear combinations of these values 
with coefficients from a prime field
$$ F_j=\sum L_k(\gamma_i), \quad j \in [0,n-1].$$

The preliminary step is the cyclotomic decomposition.

The polynomial 
$f(x)=\displaystyle\sum_{i=0}^{n-1}f_ix^i, \quad f_i\in GF(2^m)$,
can be decomposed as 
$$
f(x)=\sum_{k=0}^{l-1}L_k(x^{c_k}),\quad
L_k(y)=\sum_{j=0}^{m_k-1}f_{c_k2^j\bmod n}y^{2^j},
$$
where $\alpha^{c_k}$ is a generator of the $k$th binary conjugacy class.

Note, that the term $f_0$ can be represented as $L_0(x^0)$, where $L_0(y)=f_0 y$.

Consider the derivation of the cyclotomic algorithm.
We have 
$$F_j=f(\alpha^j)=\sum_{k=0}^{l-1}L_k(\alpha^{j\,c_k}), \quad
\alpha^{c_k} \in GF(2^{m_k}), \quad m_k \mid m.$$

All elements $(\alpha^{c_k})^j \in GF(2^{m_k})$ can be decomposed with respect to some normal basis
$\left(\gamma_k^{2^0}, \gamma_k^{2^1}, \gamma_k^{2^2}, \ldots, \gamma_k^{2^{m_k-1}}\right)$
of the subfield $GF(2^{m_k})$:
$$\displaystyle \alpha^{j\,c_k}=\displaystyle\sum_{s=0}^{m_k-1}a_{kjs} \gamma_k^{2^s}, \quad a_{kjs}\in GF(2).$$

Further, 
$$F_j=\sum_{k=0}^{l-1}L_k \left( \sum_{s=0}^{m_k-1}a_{kjs} \gamma_k^{2^s}\right) = 
\sum_{k=0}^{l-1} \sum_{s=0}^{m_k-1} a_{kjs} L_k\left(\gamma_k^{2^s}\right).$$

Each of the linearized polynomials can be evaluated at the basis points of the corresponding subfield
by the formula
$$
L_k\left(\gamma_k^{2^s}\right) = \sum_{p=0}^{m_k-1} \gamma_k^{2^{s+p}} f_{c_k 2^p},
\quad s \in [0,m_k-1], \quad k \in [0,l-1].
$$
Components of the Fourier transform of a polynomial $f(x)$ are linear
combinations of these values:
$$
F_j=f(\alpha^j)=\sum_{k=0}^{l-1}\sum_{s=0}^{m_k-1}a_{kjs}L_k\left(\gamma_k^{2^{s}}\right)
$$
$$
= \sum_{k=0}^{l-1}\sum_{s=0}^{m_k-1}a_{kjs}\left(\sum_{p=0}^{m_k-1}
\gamma_k^{2^{s+p}} f_{c_k 2^p}\right), \quad j \in [0,n-1].
$$

In matrix form, the cyclotomic algorithm can be written as
\begin{equation}
F = AL(\pi f),
\end{equation}
where $A$ is an $n \times n$ binary matrix, 
$\pi f$ is a permutation of the initial vector $f$, 
and $L$ is an $n \times n$ block diagonal matrix
\begin{equation}
L=\begin{pmatrix}
L_0 &     &        &         \\
    & L_1 &        &         \\
    &     & \ddots &         \\
    &     &        & L_{l-1} \\
\end{pmatrix}
\end{equation}
composed of basis circulants $L_k$, \, $k \in [0,l-1]$.

Multiplication of the block diagonal matrix $L$ by the vector $\pi f$ 
reduces to the computation of $l$ normalized cyclic convolutions of small lengths 
$m_k$, \, $k \in [0,l-1]$.

\subsection{The relation between the multipoint evaluation and normalized cyclic convolution}

An $m$-point cyclic convolution is $a(x) = b(x) c(x) \mod x^m-1$.
A normalized cyclic convolution of length $m$ is
$a(x) = b(x) c(x) \mod x^m-1$, if \, $\sum\limits_{i=0}^{m-1} b_i = 1$ \cite{Fedorenko11}.
The computation of a normalized cyclic convolution can be represented as 
a multiplication by a basis circulant matrix $L_k$.

The multipoint evaluation for the polynomial $t(x)$ and the point set 
$\{\epsilon_j \mid j \in J\}$, is a computation $t(\epsilon_j)$, $j \in J$ 
\cite[Chapter 10.1]{Gathen99}.
The computation of a multipoint evaluation for the point set 
$\{\alpha_1, \alpha_2, \alpha_3, \ldots, \alpha_t\}$
can be represented as a multiplication by a Vandermonde matrix.
If the point set is the binary conjugacy class, then the computation of a multipoint evaluation 
can be represented as a multiplication by a Moore--Vandermonde matrix $V_k$.

Consider the simple construction of a polynomial basis.

\begin{Lemma}[\cite{Fedorenko11}]
Let $\beta$ be an arbitrary nonzero element in $GF(2^m)$ 
whose minimal polynomial has degree $m_k$.
Then  
$\left(\beta^0, \beta^1, \beta^2, \ldots, \beta^{m_k-1}\right)$ 
is a polynomial basis for the subfield $GF(2^{m_k}) \subset GF(2^m)$.
\end{Lemma}

\begin{IEEEproof}
The proof is similar to the proof from \cite[Chapter 4, Property (M4)]{MacWilliams}.
The proof is by reductio ad absurdum.
Consider the nonzero polynomial $\sum_{i=0}^{m_k-1} a_i x^i$, $a_i \in GF(2)$, 
having $\beta$ as a root. That is, $\left(\beta^0, \beta^1, \beta^2, \ldots, \beta^{m_k-1}\right)$ 
are linearly dependent and the minimal polynomial of $\beta$ has degree less than $m_k$, a contradiction.
Therefore $\left(\beta^0, \beta^1, \beta^2, \ldots, \beta^{m_k-1}\right)$ 
are linearly independent.
\end{IEEEproof}

Let $\left((\alpha^{c_k})^0, (\alpha^{c_k})^1, (\alpha^{c_k})^2, \ldots, (\alpha^{c_k})^{m_k-1}\right)$ 
be a polynomial basis, while \\
$\left(\gamma_k^{2^0}, \gamma_k^{2^1}, \gamma_k^{2^2}, \ldots, \gamma_k^{2^{m_k-1}}\right)$
is a normal basis for the subfield $GF(2^{m_k}) \subset GF(2^m)$.

Let us construct the basis transformation matrix $M_k$ as follows:
\begin{equation}
\begin{pmatrix}
(\alpha^{c_k})^0       \\
(\alpha^{c_k})^1       \\
(\alpha^{c_k})^2       \\
\cdots                 \\
(\alpha^{c_k})^{m_k-1} \\
\end{pmatrix}
=
M_k
\begin{pmatrix}
\gamma_k^{2^0}       \\
\gamma_k^{2^1}       \\
\gamma_k^{2^2}       \\
\cdots               \\
\gamma_k^{2^{m_k-1}} \\
\end{pmatrix}.
\end{equation}
Note that the $m_k \times m_k$ matrix $M_k$ is binary and nonsingular.

\begin{Lemma}[\cite{Fedorenko11}]
The relation between the Moore--Vandermonde matrix $V_k$ and the basis circulant $L_k$ 
is $V_k^T = M_k L_k$, where $M_k$ is the basis transformation matrix.
\end{Lemma}

\begin{IEEEproof}
Let $V_k = \left(\alpha^{c_k j 2^i}\right)$, \, $i,j \in [0,m_k-1]$, \, $k \in [0,l-1]$, 
be a Moore--Vandermonde matrix. 
$L_k = \left(\gamma^{2^{i+j}}\right)$, \, $i,j \in [0,m_k-1]$, \, $k \in [0,l-1]$, 
is a basis circulant matrix.

Using (5), we have 
               
$$
\begin{pmatrix}
\begin{tabular}{llcl} 
$(\alpha^{c_k})^{0 \, 2^0}$           & $(\alpha^{c_k})^{0\, 2^1}$            & $\cdots$ & $(\alpha^{c_k})^{0\, 2^{m_k-1}}$           \\
$(\alpha^{c_k})^{1 \, 2^0}$           & $(\alpha^{c_k})^{1\, 2^1}$            & $\cdots$ & $(\alpha^{c_k})^{1\, 2^{m_k-1}}$           \\
$\cdots$                                             & $\cdots$                                             & $\cdots$ & $\cdots$                                                        \\
$(\alpha^{c_k})^{(m_k-1) \, 2^0}$ & $(\alpha^{c_k})^{(m_k-1)\, 2^1}$ & $\cdots$ & $(\alpha^{c_k})^{(m_k-1)\, 2^{m_k-1}}$ \\
\end{tabular}                                
\end{pmatrix}
$$
$$
= M_k 
\begin{pmatrix}
\gamma_k^{2^0}      &\gamma_k^{2^1} &\ldots&\gamma_k^{2^{m_k-1}} \\
\gamma_k^{2^1}      &\gamma_k^{2^2} &\ldots&\gamma_k^{2^0}       \\
\ldots            &\ldots       &\ldots&\ldots             \\
\gamma_k^{2^{m_k-1}}&\gamma_k^{2^0} &\dots &\gamma_k^{2^{m_k-2}}
\end{pmatrix}
$$
and  $V_k^T = M_k L_k$.
\end{IEEEproof}

\begin{Lemma}[\cite{Fedorenko11}]
For $m_1 = m_k$, the relation between two Moore--Vandermonde matrices is
$V_k = V_1 (M_1^T)^{-1} M_k^T$, 
where $M_1$ and $M_k$ are the basis transformation matrices.
\end{Lemma}

\begin{IEEEproof}
Using $V_k^T = M_k L_k$, we obtain $V_k = L_k^T M_k^T = L_k M_k^T$, \, 
$L_k = V_k (M_k^T)^{-1}$. For $m_1 = m_k$, we get $L_1 = L_k$ and 

$$V_k = L_k M_k^T = L_1 M_k^T = V_1 (M_1^T)^{-1} M_k^T.$$
\end{IEEEproof}

Thus the multiplicative complexity of a normalized cyclic convolution and 
a multipoint evaluation computation is the same.
They differ only by preadditions matrix.

\subsection{The relation between the Goertzel--Blahut and cyclotomic algorithms}

We introduce the following Theorem.

\begin{Theorem}[\cite{Fedorenko11}]
The multiplicative complexity of the Goertzel--Blahut algorithm 
and the cyclotomic algorithm is the same.
\end{Theorem}

\begin{IEEEproof}
The block diagonal matrix $V$ (2) 
from the Goertzel--Blahut algorithm (1) 
is factorized into a product of 
the block diagonal matrix $L$ (4) 
from the cyclotomic algorithm (3) 
and a binary block diagonal matrix.

Using $V_k = L_k M_k^T$, (2), and (4), we have 
$$
V = L 
\begin{pmatrix}
M_0^T &       &        &           \\
      & M_1^T &        &           \\
      &       & \ddots &           \\
      &       &        & M_{l-1}^T \\
\end{pmatrix}.
$$

This means that the multiplicative complexity of both algorithms is the same.
\end{IEEEproof}

\subsection{The modification of the Goertzel--Blahut algorithm}

We introduce the modification of the Goertzel--Blahut algorithm.
The algorithm is valid for each composite extension degree $m$.
Without loss of generality, we can assume that $m$ is even and $m > 2$.

The first step is a long division of $f(x)$ by each minimal polynomial $M_k(x)$, \, $k \in [0,l-1]$, over $GF(2)$:
$$
\begin{cases}
f(x) = M_k(x) q_k(x) + r_k(x), \quad   
\deg r_k(x) < \deg M_k(x) = m_k, \\
k \in [0,l-1], 
\end{cases}
$$
where $r_k(x) = \displaystyle\sum_{j=0}^{m_k-1} r_{j,k} x^j$, \,
$l$ is the number of binary conjugacy classes.

The second step for even $m$, \, $m>2$, is a long division of each $r_k(x)$, \, $k \in [0,l-1]$, 
by each minimal polynomial $M_{k,p}(x)$, \, $p \in [0,l_k -1]$, over $GF(2^{m/2})$:
$$
\begin{cases}
r_k(x) = M_{k,p}(x) q_{k,p}(x) + s_{k,p}(x), \quad   
\deg s_{k,p}(x) < \deg M_{k,p}(x), \\
k \in [0,l-1], \, p \in [0,l_k -1],
\end{cases}
$$
where $s_{k,p}(x) = \displaystyle\sum_{j=0}^{\deg M_{k,p}(x)-1} s_{j,k,p} x^j$, 
$l_k$ is the number of $2^{m/2}$-ary conjugacy classes inside the $k$th binary conjugacy class.

The third step is to evaluate the remainder at each element of the finite field:
$$
\begin{cases}
F_i = f(\alpha^i) = r_k(\alpha^i) = s_{k,p}(\alpha^i) =  \displaystyle\sum_{j=0}^{\deg M_{k,p}(x)-1} s_{j,k,p} \alpha^{ij}, \\
i \in [0,n-1], 
\end{cases}
$$
where an element $\alpha^i$ is a root of both minimal polynomials $M_k(x)$  and $M_{k,p}(x)$.

The second and third steps of this algorithm can be used for the computation of a multipoint evaluation.

\section{The multipoint evaluation for the binary conjugacy class}

Without loss of generality, we can assume that 
the binary conjugacy classes of $GF(2^m)$ have cardinality $m$.
The binary conjugacy classes of cardinality $m_i < m$ are considered
for the subfield $GF(2^{m_k}) \subset GF(2^m)$.

Consider the multipoint evaluation for the polynomial  $t(x) = \sum_{i=0}^{m-1} t_i x^i$ 
and the binary conjugacy class  $\left(\alpha^{c_k},\alpha^{c_k2},\alpha^{c_k2^2}, \ldots, \alpha^{c_k2^{m-1}}\right)$ 
of $GF(2^m)$. We compute  
$$T_i = t \left(\alpha^{c_k 2^i}\right), \, i \in [0,m-1].$$

Let us write the multipoint evaluation in matrix form: 

%\begin{tiny}
$$
\begin{pmatrix}
\begin{tabular}{l} 
$t\left(\alpha^{c_k2^0}\right)$     \\
$t\left(\alpha^{c_k2^1}\right)$     \\   
$t\left(\alpha^{c_k2^2}\right)$     \\   
$\cdots$                            \\
$t\left(\alpha^{c_k2^{m-1}}\right)$ \\
\end{tabular}
\end{pmatrix}
=
\begin{pmatrix}
\begin{tabular}{lllcl} 
$\alpha^{c_k \, 0 \, 2^0}$     & $\alpha^{c_k \, 1\, 2^0}$     & $\alpha^{c_k \, 2\, 2^0}$     & $\cdots$ & $\alpha^{c_k \, (m-1)\, 2^0}$     \\
$\alpha^{c_k \, 0 \, 2^1}$     & $\alpha^{c_k \, 1\, 2^1}$     & $\alpha^{c_k \, 2\, 2^1}$     & $\cdots$ & $\alpha^{c_k \, (m-1)\, 2^1}$     \\
$\alpha^{c_k \, 0 \, 2^2}$     & $\alpha^{c_k \, 1\, 2^2}$     & $\alpha^{c_k \, 2\, 2^2}$     & $\cdots$ & $\alpha^{c_k \, (m-1)\, 2^2}$     \\
$\cdots$                       & $\cdots$                      & $\cdots$                      & $\cdots$ & $\cdots$                          \\
$\alpha^{c_k \, 0 \, 2^{m-1}}$ & $\alpha^{c_k \, 1\, 2^{m-1}}$ & $\alpha^{c_k \, 2\, 2^{m-1}}$ & $\cdots$ & $\alpha^{c_k \, (m-1)\, 2^{m-1}}$ \\
\end{tabular}                                
\end{pmatrix}
\begin{pmatrix}
\begin{tabular}{l} 
$t_{0}$   \\
$t_{1}$   \\
$t_{2}$   \\
$\cdots$  \\
$t_{m-1}$ \\
\end{tabular}                                
\end{pmatrix},
$$
%\end{tiny}
where
$$T = V_k t,$$ 
$T = (T_i)$, $t =(t_i)$, \, $i \in [0,m-1]$, \,
$V_k = \left(\alpha^{c_k j 2^i}\right)$, \, $i,j \in [0,m-1]$, is a Moore--Vandermonde matrix.

Let us formulate the main result for the computation of a multipoint evaluation. 

\begin{Theorem}
For any finite field $GF(2^m)$ with even $m$ exists the binary conjugacy class of cardinality $m$,
the multipoint evaluation for this class reduces to two multipoint evaluations for the subfield $GF(2^{m/2})$
and $m/2$ extra multiplications.
\end{Theorem}

The multipoint evaluation matrix is shown in formula (9), 
and the number of multiplications is shown in formula (10).

In the rest of this section we consider the proof of Theorem 2.

\subsection{The properties of the binary conjugacy class} 

\begin{Lemma}
Let $\alpha^{c_k}$ be a generator of the $k$th binary conjugacy class \\
$\left(\alpha^{c_k},\alpha^{c_k2},\alpha^{c_k2^2}, \ldots,\alpha^{c_k2^{m-1}}\right)$
of cardinality $m$, $m$ is even, 
$\alpha$ is a primitive element of the field $GF(2^{m})$,
and $GF(2^{m/2})[x]$ is a polynomial ring (the ring of polynomials over field $GF(2^{m/2})$).
The polynomials 
$$\left(x-\alpha^{c_k 2^i}\right)\left(x-\alpha^{c_k 2^{m/2+i}}\right) \in GF(2^{m/2})[x], \quad i \in [0,m/2-1],$$
are the different, irreducible, and minimal polynomials over $GF(2^{m/2})$.
\end{Lemma}

\begin{IEEEproof}
Let $M_{k,i}(x) = \left(x-\alpha^{c_k 2^i}\right)\left(x-\alpha^{c_k 2^{m/2+i}}\right)$ \\
$=x^2+\left(\alpha^{c_k 2^i}+ \left(\alpha^{c_k 2^i}\right)^{2^{m/2}}\right)x+\alpha^{c_k 2^i(2^{m/2}+1)}
= x^2+\left(\beta +\beta^{2^{m/2}}\right)x+\delta^{2^i}$, 
where $\beta_i=\alpha^{c_k 2^i}$,  \, $i \in [0,m/2-1]$, 
$\delta = \left(\alpha^{2^{m/2}+1}\right)^{c_k} \in GF(2^{m/2})$. 

The element $\left(\beta_i +\beta_i^{2^{m/2}}\right)$ is a root of the polynomial $x^{2^{m/2}}+x$, hence 
$\left(\beta_i +\beta_i^{2^{m/2}}\right) \in GF(2^{m/2})$. 

Thus the polynomials $M_{k,i}(x) \in GF(2^{m/2})[x], \quad i \in [0,m/2-1],$ are the minimal and irreducible
polynomials over $GF(2^{m/2})$.

Using $M_{k}(x)=\prod_{i=0}^{m-1} \left(x-\alpha^{c_k 2^i}\right) = 
\prod_{i=0}^{m/2-1} M_{k,i}(x)$, we see that all the polynomials $M_{k,i}(x)$ are different.
\end{IEEEproof}

Let us formulate the condition that the minimal polynomial $M_{k,i}(x)$ over $GF(2^{m/2})$  
has only one coefficient in $GF(2^{m/2}) \backslash GF(2)$.

\begin{Lemma}
If $\alpha^{c_k} = \alpha^{c_k 2^{m/2}} +1$, $m$ is even, for the binary conjugacy class \\
$\left(\alpha^{c_k},\alpha^{c_k2},\alpha^{c_k2^2}, \ldots,\alpha^{c_k2^{m-1}}\right)$, then 

1) $\alpha^{c_k}$ is a root of the polynomial $x^{2^{m/2}}+x+1$;

2) $\alpha^{c_k 2^i} = \alpha^{c_k 2^{m/2+i}} +1$ for all $i \in [0,m/2-1]$;

3) the minimal polynomials over $GF(2^{m/2})$ are 
$$M_{k,i}(x)= \left(x-\alpha^{c_k 2^i}\right)\left(x-\alpha^{c_k 2^{m/2+i}}\right) = 
x^2+x+\alpha^{c_k\left(2^i + 2^{m/2+i}\right)} = 
x^2+x+\delta^{2^i},$$
where $\delta = \left(\alpha^{2^{m/2}+1}\right)^{c_k} \in GF(2^{m/2})$, 
$i \in [0,m/2-1]$;

4) the minimal polynomials $M_{k,i}(x)$, \, $i \in [0,m/2-1]$, over $GF(2^{m/2})$ are different.
The elements $\delta^{2^i}$ , \, $i \in [0,m/2-1]$ are different, too;

5) the minimal polynomial over $GF(2)$ is 
$$M_{k}(x)=\prod_{i=0}^{m-1} \left(x-\alpha^{c_k 2^i}\right) = 
\prod_{i=0}^{m/2-1} M_{k,i}(x)=\prod_{i=0}^{m/2-1} \left(x^2+x+\delta^{2^i}\right).$$
\end{Lemma}

\begin{IEEEproof}
The proof is trivial.
\end{IEEEproof}

\subsection{The binary conjugacy class choice} 

\begin{Theorem}
For any finite field $GF(2^m)$ with even $m$ exists the binary conjugacy class 
$\left(\alpha^{c_k},\alpha^{c_k2},\alpha^{c_k2^2}, \ldots,\alpha^{c_k2^{m-1}}\right)$
such that $\alpha^{c_k} = \alpha^{c_k 2^{m/2}} +1$. 
\end{Theorem}

To prove this Theorem, we need two Lemmas.

\begin{Lemma}
For any finite field $GF(2^m)$ with even $m$:

$$x^{2^{m/2}}+x+1 \mid x^{2^m}+x.$$
\end{Lemma}

\begin{IEEEproof}
The proof is similar to the proof from \cite[Theorems 11.34 and 11.35]{Berlekamp}.
Let $L(x)=x^{2^{m/2}}+x$, $A(x)=L(x)+1=x^{2^{m/2}}+x+1$, and $B(x)= (L(x))^{2^{m/2}-1}+1$.

We obviously have $L(x)B(x)=x^{2^m}+x$.

Let $r$ be a root of the polynomial $A(x)$. Since $L(r)=1$ and $B(r)=0$, we see that all roots of the polynomial $A(x)$
are roots of the polynomial $B(x)$. Hence, $A(x)  \mid B(x)$ and $A(x) \mid x^{2^m}+x$. 
\end{IEEEproof} 

\begin{Lemma}
For any finite field $GF(2^m)$ with even $m$ exists 
the minimal polynomial $M_k(x)$ of degree $m$ over $GF(2)$ such that

$$M_k(x) \mid x^{2^{m/2}}+x+1.$$
\end{Lemma}

\begin{IEEEproof}
Let $\Irr(i)$ be a number of irreducible binary polynomials of degree $i$.
From \cite[Theorem 3.35]{Berlekamp} it follows that $m \Irr(i) > 2^m - 2^{m/2+1}$.
The number of the finite field $GF(2^m)$ elements
belonging to the binary conjugacy classes of cardinality $m$ is $m \Irr(m)$.
The number of the finite field $GF(2^m)$ elements without the roots of the polynomials
$x^{2^{m/2}}+x$ and $x^{2^{m/2}}+x+1$ equals $2^m - 2^{m/2+1}$.
Since there are no elements belonging to the binary conjugacy class 
of cardinality $m$ within the subfield $GF(2^{m/2})$, 
the set of the polynomial $x^{2^{m/2}}+x+1$ roots contains at least
one element (and hence the binary conjugacy class), which belongs to 
the binary conjugacy class of cardinality $m$.
The minimal polynomial over $GF(2)$ for this binary conjugacy class divides the polynomial $x^{2^{m/2}}+x+1$.
\end{IEEEproof}

This completes the proof of Theorem 3.

The binary conjugacy class choice consists of two steps:

1. construct the minimal polynomial $M_k(x)$  of degree $m$ over $GF(2)$ such that

$$M_k(x) \mid x^{2^{m/2}}+x+1;$$

2. choose a root $\alpha^{c_k}$ of the minimal polynomial $M_k(x)$ 
as a generator of the $k$th  binary conjugacy class 
$\left(\alpha^{c_k},\alpha^{c_k2},\alpha^{c_k2^2}, \ldots,\alpha^{c_k2^{m-1}}\right)$.

\subsection{Two divisions levels in the multipoint evaluation} 

The second and third steps of the modification of the Goertzel--Blahut algorithm
includes the upper and lower levels of divisions in the multipoint evaluation, respectively.
From Lemma 5 it follows that the quadratic polynomial $M_{k,i}(x)$ 
has only one coefficient in $GF(2^{m/2}) \backslash GF(2)$,
and a division by this polynomial is very simple.

The upper level of divisions is a long division of polynomial $t(x) = \sum_{i=0}^{m-1} t_i x^i$  
by each quadratic polynomial 
$M_{k,i}(x) = x^2+x+\delta^{2^i}$, \, $i \in [0,m/2-1]$:
$$t(x) = \left(x^2+x+\delta^{2^i}\right) q_i(x)+u_i(x), \quad i \in [0,m/2-1],$$
where $u_i(x) = u_{i,1}x+u_{i,0}$, 
$\delta = \left(\alpha^{2^{m/2}+1}\right)^{c_k} \in GF(2^{m/2})$. 

The lower level of divisions is evaluating the remainder $u_i(x)$ at 
pair conjugates $\alpha^{c_k 2^i}$ and $\alpha^{c_k 2^{m/2+i}}$
(with respect to $GF(2^{m/2})$) for all $i \in [0,m/2-1]$:

$$
\begin{tabular}{lllll} 
$T_{i}$          &= $u_i\left(\alpha^{c_k 2^i}\right)$              &= $u_{i,1} \alpha^{c_k 2^i}$                      &+ $ u_{i,0}$ &                                 \\
$T_{m/2+i}$ &= $u_i\left(\alpha^{c_k 2^{m/2+i}}\right)$ &= $u_{i,1} \left(\alpha^{c_k 2^i}+1\right)$ &+ $u_{i,0}$  &= $T_{i} + u_{i,1}$ \\
\end{tabular}.
$$

\subsection{The matrix form of two divisions levels in the multipoint evaluation} 

We decompose the multipoint evaluation matrix into two factors

$$V_k = U_{lower} U_{upper},$$
where $U_{lower}, \, U_{upper}$ are nonsingular.

\subsubsection{The lower level of divisions}

Consider the matrix form of the lower level of divisions

$$
\begin{pmatrix}
\begin{tabular}{l} 
$T_0$      \\
$T_1$      \\
$T_2$      \\
$\cdots$   \\
$T_{m/2-1}$\\ \hline
$T_{m/2}$  \\
$T_{m/2+1}$\\
$T_{m/2+2}$\\
$\cdots$   \\
$T_{m-1}$  \\ 
\end{tabular}
\end{pmatrix}
= U_{lower}
\begin{pmatrix}
\begin{tabular}{l} 
$u_{0,0}$     \\ 
$u_{0,1}$     \\ \hline
$u_{1,0}$     \\
$u_{1,1}$     \\ \hline
$u_{2,0}$     \\
$u_{2,1}$     \\ \hline
$\cdots$      \\ 
$\cdots$      \\ \hline
$u_{m/2-1,0}$ \\ 
$u_{m/2-1,1}$ \\ 
\end{tabular}
\end{pmatrix},
$$
$$ U_{lower} $$
$$
= 
\begin{pmatrix}
\begin{tabular}{llllllllllll} 
1             & $\alpha^{c_k}$                 & 0            & 0                                             & 0             & 0                                             & 0  & $\cdots$       & 0 & 0 & 0                        \\
0             & 0                                         & 1            & $\alpha^{c_k 2}$                  & 0             & 0                                             & 0  & $\cdots$       & 0 & 0 & 0                        \\
0             & 0                                         & 0            & 0                                             & 1             & $\alpha^{c_k 2^2}$             & 0  & $\cdots$       & 0 &0 & 0                        \\
$\cdots$ & $\cdots$                             & $\cdots$ & $\cdots$                                 & $\cdots$ & $\cdots$                                 & $\cdots$ & $\cdots$ & $\cdots$ & $\cdots$ & $\cdots$                 \\
0             & 0                                         & 0            & 0                                             & 0             & 0                                             & 0  & $\cdots$ & 0        & 1        & $\alpha^{c_k 2^{m/2-1}}$ \\ \hline
1             & $\alpha^{c_k 2^{m/2}}$ & 0            & 0                                             & 0             & 0                                              & 0     & $\cdots$         & 0 & 0 & 0                        \\
0             & 0                                         & 1            & $\alpha^{c_k 2^{m/2+1}}$ & 0            & 0                                              & 0     & $\cdots$         &0 & 0 & 0                        \\
0             & 0                                         & 0            & 0                                             & 1             & $\alpha^{c_k 2^{m/2+2}}$ & 0     & $\cdots$         & 0 & 0 & 0                        \\
$\cdots$ & $\cdots$                             & $\cdots$ & $\cdots$                                 & $\cdots$ & $\cdots$                                  & $\cdots$ & $\cdots$ & $\cdots$ & $\cdots$ & $\cdots$                 \\
0             & 0                                        & 0             & 0                                             & 0             & 0                                              & 0 & $\cdots$            & 0        & 1        & $\alpha^{c_k 2^{m-1}}$   \\ 
\end{tabular}                                
\end{pmatrix} 
$$
$$
=
\begin{pmatrix}
\begin{tabular}{l|l} 
 $I_{m/2}$ & O               \\ \hline
 $I_{m/2}$ & $I_{m/2}$ \\
\end{tabular}                                
\end{pmatrix} 
\begin{pmatrix}
\begin{tabular}{c|c} 
 $I_{m/2}$ & 
{$
\begin{matrix}
\begin{tabular}{lllll} 
$\alpha^{c_k}$  &                                &                                     &                  &                                               \\
                            & $\alpha^{c_k 2}$  &                                     &                  &                                               \\
                            &                                & $\alpha^{c_k 2^2}$  &                  &                                                \\ 
                            &                                &                                     &  $\ddots$  &                                                \\ 
                            &                                &                                     &                  &  $\alpha^{c_k 2^{m/2-1}}$ \\ 
\end{tabular}                                
\end{matrix}
$}
               \\ \hline
 O            & $I_{m/2}$ \\
\end{tabular}                                
\end{pmatrix} 
\Pi,
$$
where $I_{m/2}$ is the $(m/2) \times (m/2)$ identity matrix, 
O is the $(m/2) \times (m/2)$ all-zero matrix, \\ 
the permutation matrix is
$$
\Pi = 
\begin{pmatrix}
\begin{tabular}{ccccccccccc} 
1 & 0 & 0 & 0 & 0 & 0 & 0 & $\cdots$ & 0 & 0 & 0 \\
0 & 0 & 1 & 0 & 0 & 0 & 0 & $\cdots$ & 0 & 0 & 0 \\
0 & 0 & 0 & 0 & 1 & 0 & 0 & $\cdots$ & 0 & 0 & 0 \\ 
$\cdots$  & $\cdots$ & $\cdots$ & $\cdots$ & $\cdots$ & $\cdots$ & $\cdots$ & $\cdots$ & $\cdots$ & $\cdots$ & $\cdots$ \\
0 & 0 & 0 & 0 & 0 & 0 & 0 & $\cdots$ & 0 & 1 & 0 \\ \hline
0 & 1 & 0 & 0 & 0 & 0 & 0 & $\cdots$ & 0 & 0 & 0 \\
0 & 0 & 0 & 1 & 0 & 0 & 0 & $\cdots$ & 0 & 0 & 0 \\
0 & 0 & 0 & 0 & 0 & 1 & 0 & $\cdots$ & 0 & 0 & 0 \\ 
$\cdots$  & $\cdots$ & $\cdots$ & $\cdots$ & $\cdots$ & $\cdots$ & $\cdots$ & $\cdots$ & $\cdots$ & $\cdots$ & $\cdots$ \\
0 & 0 & 0 & 0 & 0 & 0 & 0 & $\cdots$ & 0 & 0 & 1 \\
\end{tabular}                                
\end{pmatrix}, 
$$
or  $\Pi= (\Pi_{ij}), \, i,j \in [0,m-1]$: 
\begin{equation}
\Pi_{ij} = 
\begin{cases}
1, \hbox{ if } (j=2i) \hbox{ and } i \in [0,m/2-1], \\
1, \hbox{ if } (j=2i+1-m) \hbox{ and } i \in [m/2,m-1], \\
0, \hbox{ otherwise. }\\
\end{cases}
\end{equation}

\subsubsection{The upper level of divisions}

Consider the matrix form of the upper level of divisions for $m > 2$

$$
\begin{pmatrix}
\begin{tabular}{l} 
$u_{0,0}$     \\ 
$u_{0,1}$     \\ \hline
$u_{1,0}$     \\
$u_{1,1}$     \\ \hline
$u_{2,0}$     \\
$u_{2,1}$     \\ \hline
$\cdots$      \\ 
$\cdots$      \\ \hline
$u_{m/2-1,0}$ \\ 
$u_{m/2-1,1}$ \\ 
\end{tabular}
\end{pmatrix} 
= U_{upper}
\begin{pmatrix}
\begin{tabular}{l} 
$t_{0}$   \\
$t_{1}$   \\
$t_{2}$   \\
$\cdots$  \\
$t_{m-1}$ \\
\end{tabular}                                
\end{pmatrix}.
$$

\begin{Theorem}
The matrix $U_{upper}$ can be represented as
$$
U_{upper} = 
\Pi^{-1}
\begin{pmatrix}
\begin{tabular}{l|l} 
$\Delta$ & O             \\ \hline
O             & $\Delta$ \\
\end{tabular}                                
\end{pmatrix}
\Pi B^{-1},
$$
where $m$ is even, $m > 2$, 
the permutation matrix $\Pi$ is defined in formula (6), 
$$
\Delta = 
\begin{pmatrix}
\begin{tabular}{lllcl} 
$\delta^{0 \, 2^0}$     & $\delta^{1\, 2^0}$     & $\delta^{2\, 2^0}$     & $\cdots$ & $\delta^{(m/2-1)\, 2^0}$     \\
$\delta^{0 \, 2^1}$     & $\delta^{1\, 2^1}$     & $\delta^{2\, 2^1}$     & $\cdots$ & $\delta^{(m/2-1)\, 2^1}$     \\
$\delta^{0 \, 2^2}$     & $\delta^{1\, 2^2}$     & $\delta^{2\, 2^2}$     & $\cdots$ & $\delta^{(m/2-1)\, 2^2}$     \\
$\cdots$                       & $\cdots$                      & $\cdots$                      & $\cdots$ & $\cdots$                          \\
$\delta^{0 \, 2^{m/2-1}}$ & $\delta^{1\, 2^{m/2-1}}$ & $\delta^{2\, 2^{m/2-1}}$ & $\cdots$ & $\delta^{(m/2-1)\, 2^{m/2-1}}$ \\
\end{tabular}                                
\end{pmatrix}
$$
$$=\left(\delta^{j 2^i}\right), \quad i,j \in [0,m/2-1],$$
is a Moore--Vandermonde matrix, 
$\delta = \left(\alpha^{2^{m/2}+1}\right)^{c_k} \in GF(2^{m/2})$, 
O is the $(m/2) \times (m/2)$ all-zero matrix,  
the binary matrix $B$ is defined in Lemma 10.
\end{Theorem}

To prove this Theorem, we need several Lemmas.

\begin{Lemma}
Polynomial $r_1x+r_0$  is the remainder polynomial of the polynomial $t(x) = \sum_{i=0}^{m-1} t_i x^i$, \,
$m>2$, when divided by the quadratic polynomial $x^2+x+\epsilon$, \, $\epsilon \in GF(2^m)$, in characteristic two finite field.
Then 

$$
\begin{pmatrix}
\begin{tabular}{l} 
$r_0$     \\ \hline
$r_1$     \\ 
\end{tabular}
\end{pmatrix} 
= U
\begin{pmatrix}
\begin{tabular}{l} 
$t_{0}$   \\
$t_{1}$   \\
$t_{2}$   \\
$\cdots$  \\
$t_{m-1}$ \\
\end{tabular}                                
\end{pmatrix},
$$
where the remainder matrix is
$$
U=
\begin{pmatrix}
\begin{tabular}{lllll} 
$r_{0,0}$ & $r_{0,1}$ & $r_{0,2}$ & $\cdots$ & $r_{0,m-1}$ \\ \hline
$r_{1,0}$ & $r_{1,1}$ & $r_{1,2}$ & $\cdots$ & $r_{1,m-1}$ \\
\end{tabular}                                
\end{pmatrix},
$$
$$
r_{j,i} = r_{j,i-1}+\epsilon r_{j,i-2}, \quad j \in [0,1], \quad i \ge 2,
$$
with initial conditions
$$
\begin{cases}
r_{1,0} =0, \\
r_{1,1} =1, \\
r_{0,0} =1, \\
r_{0,1} =0. \\
\end{cases}
$$
\end{Lemma}

\begin{IEEEproof}
Let  $\lambda$ and $\mu$ be roots of the polynomial $x^2+x+\epsilon$. \\
We have $x^2+x+\epsilon=(x-\lambda)(x-\mu)=x^2+(\lambda+\mu)x+\lambda \mu$
and 
$
\begin{cases}
\lambda+\mu =1  \\
\lambda \mu = \epsilon \\
\end{cases}
.$

From $t(x)= (x^2+x+\epsilon) q(x) + (r_1x+r_0)$  it follows that
$
\begin{cases}
t(\lambda) = r_1 \lambda + r_0 \\
t(\mu) = r_1 \mu + r_0 \\
\end{cases} 
\Rightarrow 
$

$
\begin{cases}
r_1=t(\lambda)+t(\mu) = \sum_{i=0}^{m-1} t_i \lambda^i +\sum_{i=0}^{m-1} t_i \mu^i \\
r_0= t(\lambda)+ r_1 \lambda = t(\lambda) \mu + t(\mu) \lambda = \sum_{i=0}^{m-1} t_i (\lambda^i \mu + \mu^i \lambda) \\
\end{cases} 
\Rightarrow 
$

\smallskip

$
\begin{cases}
r_1= \sum_{i=0}^{m-1} t_i (\lambda^i +\mu^i) = \sum_{i=0}^{m-1} t_i r_{1,i} \\ 
r_0=t_0+\sum_{i=1}^{m-1} t_i \epsilon (\lambda^{i-1} + \mu^{i-1}) = \sum_{i=0}^{m-1} t_i r_{0,i} \\
\end{cases}. 
$

\smallskip

Hence, we obtain

\smallskip

\begin{small}
$r_{1,i}= \lambda^i +\mu^i = 
\begin{cases}
(\lambda^{i-1}+\mu^{i-1})(\lambda+\mu)-\lambda\mu(\lambda^{i-2}+\mu^{i-2}) = r_{1,i-1}+\epsilon r_{1,i-2}, & \hbox{ if } i \in [2,m-1], \\
0, & \hbox{ if } i=0,\\
1, & \hbox{ if } i=1,\\
\end{cases} 
$
\end{small}
\\
that is, 

\begin{equation}
r_{1,i}=r_{1,i-1}+\epsilon r_{1,i-2}, \quad i \ge 2.  
\end{equation}

Further, we obtain
$r_{0,i} = 
\begin{cases}
\epsilon (\lambda^{i-1}+\mu^{i-1}) = \epsilon r_{1,i-1}, & \hbox{ if } i \in [1,m-1], \\
1, & \hbox{ if } i=0,\\
\end{cases} 
$ \\
that is, 

\begin{equation}
r_{0,i} =  \epsilon r_{1,i-1}, \quad i \ge 1. 
\end{equation}

Using (7) and (8), we get

$$r_{0,i} = r_{0,i-1}+\epsilon r_{0,i-2}, \quad  i \ge 2.$$
\end{IEEEproof}

\begin{Corollary}
If $m$ is even and 
$\epsilon = \delta = \delta^{2^0} = \left(\alpha^{2^{m/2}+1}\right)^{c_k} \in GF(2^{m/2})$, 
then all elements of the remainder matrix

$$
U=U(0)=
\begin{pmatrix}
\begin{tabular}{lllll} 
$r_{0,0}$ & $r_{0,1}$ & $r_{0,2}$ & $\cdots$ & $r_{0,m-1}$ \\ \hline
$r_{1,0}$ & $r_{1,1}$ & $r_{1,2}$ & $\cdots$ & $r_{1,m-1}$ \\
\end{tabular}                                
\end{pmatrix}
$$
belong to the subfield $GF(2^{m/2}) \subset GF(2^m)$.
\end{Corollary}

\begin{Corollary}
If $m$ is even and $\epsilon = \delta^{2^i}$, then the remainder matrix is
$U(i)=[U(0)]^{2^i}=[U]^{2^i}$,
where matrix $[U]^j$ consists of $j$th degree of all elements of the matrix $U$.
\end{Corollary}

\begin{Lemma}
Let us $\delta = \left(\alpha^{2^{m/2}+1}\right)^{c_k} \in GF(2^{m/2})$, 
$m$ is even,  $m > 2$, \\
then $\left(\delta^{0},\delta^{1},\delta^{2}, \ldots,\delta^{m/2-1}\right)$
is a polynomial basis for the subfield $GF(2^{m/2}) \subset GF(2^m)$.
\end{Lemma}

\begin{IEEEproof}
Consider a $(m/2) \times (m/2)$ Moore--Vandermonde matrix  
$\Delta =  \left(\delta^{j 2^i}\right), \, i,j \in [0,m/2-1]$.
Since all elements $\delta^{2^i}$ , \, $i \in [0,m/2-1]$, are different (see Lemma 5, Property 4), 
it follows that the Vandermonde matrix  $\Delta$ is nonsingular. 
The Moore matrix  $\Delta$ is nonsingular if and only if  
$\left(\delta^{0},\delta^{1},\delta^{2}, \ldots,\delta^{m/2-1}\right)$ 
is a basis \cite[Chapter 4, Lemma 18]{MacWilliams}.
\end{IEEEproof}

\begin{Lemma}
If the remainder matrix $U$ is defined in Lemma 8 for $\epsilon = \delta$
$$
U=U(0)=
\begin{pmatrix}
\begin{tabular}{lllll} 
$r_{0,0}$ & $r_{0,1}$ & $r_{0,2}$ & $\cdots$ & $r_{0,m-1}$ \\ \hline
$r_{1,0}$ & $r_{1,1}$ & $r_{1,2}$ & $\cdots$ & $r_{1,m-1}$ \\
\end{tabular}                                
\end{pmatrix},
$$
$m$ is even, $m > 2$, \, 
$\left(\delta^{0},\delta^{1},\delta^{2}, \ldots,\delta^{m/2-1}\right)$
is a polynomial basis for the subfield $GF(2^{m/2}) \subset GF(2^m)$, 
and $\delta = \left(\alpha^{2^{m/2}+1}\right)^{c_k} \in GF(2^{m/2})$, 
then there is the nonsingular binary matrix $B$ such that

$$
U B = 
\begin{pmatrix}
\begin{tabular}{l|l|l|l|l|l|l|l|l} 
$\delta^{0}$ & 0            & $\delta^{1}$ & 0            & $\delta^{2}$ & 0            & $\cdots$ & $\delta^{m/2-1}$ & 0                \\ \hline
0            & $\delta^{0}$ & 0            & $\delta^{1}$ & 0            & $\delta^{2}$ & $\cdots$ & 0                & $\delta^{m/2-1}$ \\
\end{tabular}                                
\end{pmatrix}.
$$
\end{Lemma}

\begin{IEEEproof}
Let the matrix $U_j$ consist of the first $2(j+1)$ columns of the matrix $U$
$$
U_j =
\begin{pmatrix}
\begin{tabular}{llll|ll} 
$r_{0,0}$ & $r_{0,1}$ & $r_{0,2}$ & $\cdots$ & $r_{0,2j}$ & $r_{0,2j+1}$\\ \hline
$r_{1,0}$ & $r_{1,1}$ & $r_{1,2}$ & $\cdots$ & $r_{1,2j}$ & $r_{1,2j+1}$\\
\end{tabular}                                
\end{pmatrix}.
$$

We use mathematical induction on $j$.

1. The basis. For $j=1$ the matrix $U_1$ is 
$
U_1 =
\begin{pmatrix}
\begin{tabular}{ll|ll} 
1 & 0 & $\delta$ & $\delta$     \\ \hline
0 & 1 & 1        & $\delta + 1$ \\
\end{tabular}                                
\end{pmatrix}.
$

\smallskip

2. The inductive step. For $j$ there are next properties:

a) for each of the elements $r_{0,2j}, r_{1,2j}, r_{0,2j+1}, r_{1,2j+1}$  
there are binary coefficients $a_{i,s}$, \, $i \in [0,j]$, \, $s \in [1,4]$, 
such that 

a1) $r_{0,2j} = \delta^j + \sum_{i=0}^{j-1} a_{i,1} \delta^i$,

a2) $r_{0,2j+1} = \sum_{i=0}^{j} a_{i,2} \delta^i$,

a3) $r_{1,2j} = \sum_{i=0}^{j-1} a_{i,3} \delta^i$,

a4) $r_{1,2j+1} = \delta^j + \sum_{i=0}^{j-1} a_{i,4} \delta^i$;

b) there exists the nonsingular binary matrix $B_j$ such that

$$
U_j B_j = 
\begin{pmatrix}
\begin{tabular}{l|l|l|l|l|l|l|l|l} 
$\delta^{0}$ & 0            & $\delta^{1}$ & 0            & $\delta^{2}$ & 0            & $\cdots$ & $\delta^{j}$ & 0                \\ \hline
0            & $\delta^{0}$ & 0            & $\delta^{1}$ & 0            & $\delta^{2}$ & $\cdots$ & 0                & $\delta^{j}$ \\
\end{tabular}                                
\end{pmatrix}.
$$

For $j+1$ there are:

the matrix
$
U_{j+1} =
\begin{pmatrix}
\begin{tabular}{llll|ll|ll} 
$r_{0,0}$ & $r_{0,1}$ & $r_{0,2}$ & $\cdots$ & $r_{0,2j}$ & $r_{0,2j+1}$ & $r_{0,2j+2}$ & $r_{0,2j+3}$ \\ \hline
$r_{1,0}$ & $r_{1,1}$ & $r_{1,2}$ & $\cdots$ & $r_{1,2j}$ & $r_{1,2j+1}$ & $r_{1,2j+2}$ & $r_{1,2j+3}$\\
\end{tabular}                                
\end{pmatrix};
$

\smallskip

c) for each of the elements $r_{0,2j+2}, r_{0,2j+3}, r_{1,2j+2}, r_{1,2j+3}$  
there are binary coefficients $b_{i,s}$, \, $i \in [0,j+1]$, \, $s \in [1,4]$, 
such that 

c1) using Lemma 8 and Properties (a1,a2), we get 
$r_{0,2j+2} = r_{0,2j+1} + \delta r_{0,2j} = 
\sum_{i=0}^{j} a_{i,2} \delta^i + 
\delta \left(\delta^j + \sum_{i=0}^{j-1} a_{i,1} \delta^i \right) = 
\delta^{j+1} + \sum_{i=0}^{j} b_{i,1} \delta^i$,

c2) using Lemma 8 and Properties (a2,c1), we get 
$r_{0,2j+3} = r_{0,2j+2} + \delta r_{0,2j+1} = 
\delta^{j+1} + \sum_{i=0}^{j} b_{i,1} \delta^i +
\delta \sum_{i=0}^{j} a_{i,2} \delta^i = 
\sum_{i=0}^{j+1} b_{i,2} \delta^i$,

c3) using Lemma 8 and Properties (a3,a4), we get 
$r_{1,2j+2} = r_{1,2j+1} + \delta r_{1,2j} = 
\delta^{j} + \sum_{i=0}^{j-1} a_{i,4} \delta^i +
\delta \sum_{i=0}^{j-1} a_{i,3} \delta^i = 
\sum_{i=0}^{j} b_{i,3} \delta^i$,

c4) using Lemma 8 and Properties (a4,c3), we get 
$r_{1,2j+3} = r_{1,2j+2} + \delta r_{1,2j+1} = 
\sum_{i=0}^{j} b_{i,3} \delta^i + 
\delta \left(\delta^j + \sum_{i=0}^{j-1} a_{i,4} \delta^i \right) = 
\delta^{j+1} + \sum_{i=0}^{j} b_{i,4} \delta^i$;

d) there exists the nonsingular binary matrix $B_{j+1}$ such that

$$
U_{j+1} B_{j+1} = 
\begin{pmatrix}
\begin{tabular}{l|l|l|l|l|l|l|l|l|l|l} 
$\delta^{0}$ & 0            & $\delta^{1}$ & 0            & $\delta^{2}$ & 0            & $\cdots$ & $\delta^{j}$ & 0 & $\delta^{j+1}$ & 0 \\ \hline
0            & $\delta^{0}$ & 0            & $\delta^{1}$ & 0            & $\delta^{2}$ & $\cdots$ & 0                & $\delta^{j}$ & 0 & $\delta^{j+1}$ \\ 
\end{tabular}                                
\end{pmatrix}.
$$
\end{IEEEproof}

\begin{IEEEproof}[Proof of Theorem 4]
The remainder matrix $U(i)$ is calculated in Lemma 8 for a long division of 
the polynomial $t(x) = \sum_{j=0}^{m-1} t_j x^j$, \, $m>2$, 
by the quadratic polynomial $x^2+x+\delta^{2^i}$, \, $i \in [0,m/2-1]$.

From Corollary 2 it follows that $U(i)=[U(0)]^{2^i}=[U]^{2^i}$.
Then
$$
U_{upper} = 
\begin{pmatrix}
\begin{tabular}{l} 
$U(0)$        \\ \hline
$U(1)$        \\ \hline
$U(2)$        \\ \hline
$\cdots$      \\ \hline
$U(m/2-1)$ \\ 
\end{tabular}
\end{pmatrix} 
=
\begin{pmatrix}
\begin{tabular}{l} 
$U$        \\ \hline
$[U]^2$        \\ \hline
$[U]^{2^2}$        \\ \hline
$\cdots$      \\ \hline
$[U]^{2^{m/2-1}}$ \\ 
\end{tabular}
\end{pmatrix}.
$$

Using Lemma 10, we have

$$U_{upper} B$$
\begin{tiny}
$$
=
\begin{pmatrix}
\begin{tabular}{l|l|l|l|l|l|l|l} 
$\delta^{0}$ & 0            & $\delta^{1}$ & 0            & $\delta^{2}$ & $\cdots$ & $\delta^{m/2-1}$ & 0                \\ 
0            & $\delta^{0}$ & 0            & $\delta^{1}$ & 0            & $\cdots$ & 0                & $\delta^{m/2-1}$ \\ \hline
$\delta^{0\, 2^1}$ & 0            & $\delta^{1\, 2^1}$ & 0            & $\delta^{2\, 2^1}$ & $\cdots$ & $\delta^{(m/2-1)\, 2^1}$ & 0                \\ 
0            & $\delta^{0\, 2^1}$ & 0            & $\delta^{1\, 2^1}$ & 0            & $\cdots$ & 0                & $\delta^{(m/2-1)\, 2^1}$ \\ \hline
$\delta^{0\, 2^2}$ & 0            & $\delta^{1\, 2^2}$ & 0            & $\delta^{2\, 2^2}$ & $\cdots$ & $\delta^{(m/2-1)\, 2^2}$ & 0                \\ 
0            & $\delta^{0\, 2^2}$ & 0            & $\delta^{1\, 2^2}$ & 0            & $\cdots$ & 0                & $\delta^{(m/2-1)\, 2^2}$ \\ \hline
$\cdots$ & $\cdots$ & $\cdots$ & $\cdots$ & $\cdots$  & $\cdots$ & $\cdots$ & $\cdots$ \\ \hline
$\delta^{0\, 2^{m/2-1}}$ & 0            & $\delta^{1\, 2^{m/2-1}}$ & 0            & $\delta^{2\, 2^{m/2-1}}$ &  $\cdots$ & $\delta^{(m/2-1)\, 2^{m/2-1}}$ & 0 \\ 
0            & $\delta^{0\, 2^{m/2-1}}$ & 0            & $\delta^{1\, 2^{m/2-1}}$ & 0            & $\cdots$ & 0                & $\delta^{(m/2-1)\, 2^{m/2-1}}$ \\ 
\end{tabular}                                
\end{pmatrix}
$$
\end{tiny}
$$
= \Pi^{-1}
\begin{pmatrix}
\begin{tabular}{l|l} 
$\Delta$ & O             \\ \hline
O             & $\Delta$ \\
\end{tabular}                                
\end{pmatrix}
\Pi.
$$

Finally, we obtain

$$
U_{upper} = 
\Pi^{-1}
\begin{pmatrix}
\begin{tabular}{l|l} 
$\Delta$ & O             \\ \hline
O             & $\Delta$ \\
\end{tabular}                                
\end{pmatrix}
\Pi B^{-1}.
$$
\end{IEEEproof}

\subsubsection{The multipoint evaluation matrix}

The multipoint evaluation matrix for even $m$, \, $m \ge 2$, is

$$
V_k  = U_{lower} U_{upper}
$$
\begin{small}
\begin{equation}
=
\begin{pmatrix}
\begin{tabular}{l|l} 
 $I_{m/2}$ & O               \\ \hline
 $I_{m/2}$ & $I_{m/2}$ \\
\end{tabular}                                
\end{pmatrix} 
\begin{pmatrix}
\begin{tabular}{c|c} 
 $I_{m/2}$ & 
{$
\begin{matrix}
\begin{tabular}{lllll} 
$\alpha^{c_k}$  &                                &                                     &                  &                                               \\
                            & $\alpha^{c_k 2}$  &                                     &                  &                                               \\
                            &                                & $\alpha^{c_k 2^2}$  &                  &                                                \\ 
                            &                                &                                     &  $\ddots$  &                                                \\ 
                            &                                &                                     &                  &  $\alpha^{c_k 2^{m/2-1}}$ \\ 
\end{tabular}                                
\end{matrix}
$}
               \\ \hline
 O             & $I_{m/2}$ \\
\end{tabular}                                
\end{pmatrix} 
\begin{pmatrix}
\begin{tabular}{l|l} 
$\Delta$ & O             \\ \hline
O             & $\Delta$ \\
\end{tabular}                                
\end{pmatrix}
\Pi B^{-1},
\end{equation}
\end{small}

\noindent
where 
$I_{m/2}$ is the $(m/2) \times (m/2)$ identity matrix,  
O is the $(m/2) \times (m/2)$ all-zero matrix,  
the permutation matrix $\Pi$ is defined in formula (6), 
$\Delta = \left(\delta^{j 2^i}\right)$, \, $i,j \in [0,m/2-1]$, is a Moore--Vandermonde matrix, 
$\delta = \left(\alpha^{2^{m/2}+1}\right)^{c_k} \in GF(2^{m/2})$, 
the binary matrix $B$ is defined in Lemma 10.

\subsection{The complexity of the multipoint evaluation}

The recursive formula for the number of multiplications of the multipoint evaluation is
\begin{equation}
\Mult(m) = 2 \Mult(m/2) + m/2,
\end{equation}
with initial condition $\Mult(1)=0$.

This recursion is satisfied by $\Mult(m) = \frac{1}{2} m \log_2 m$ for $m=2^i$, \, $i \ge 0$.

Let us remember (see Lemma 2) that the multiplicative complexity 
of a normalized cyclic convolution and 
a multipoint evaluation computation is the same.
The complexity of the multipoint evaluation for old methods \cite{Blahut83}, \cite[Table 1]{Wu},  
and for even $m$ of the novel method is shown in Table \ref{table1}.

\begin{table}[!t]
\renewcommand{\arraystretch}{1.3}
\caption{The complexity of the multipoint evaluation}
\label{table1}
\begin{center}
\begin{tabular}{|r||r|r|}\hline
    & \bfseries old methods     & \bfseries novel method    \\ \hline
$m$ & \bfseries multiplications & \bfseries multiplications \\ \hline\hline
 1  &        0&    0   \\
 2  &        1&    1   \\
 3  &        3&  ---   \\
 4  &        5&    4   \\
 5  &        9&  ---   \\
 6  &       10&    9  \\
 7  &       12&  ---  \\
 8  &       19&   12 \\
 9  &       18&  ---  \\
10  &       28&  23 \\
11  &       42&  --- \\
12  &       32&  24 \\ \hline
\end{tabular}                                                             
\end{center}
\end{table}

\subsection{Examples}
\subsubsection{$m=2$}

Let $c_k$=1, while the binary conjugacy class is 
$\left(\alpha^{1},\alpha^{2}\right)$ of $GF(2^2)$.

There is the lower level of divisions only.

$$
V_1 = 
\begin{pmatrix}
\begin{tabular}{ll} 
 1 & $\alpha^1$       \\ 
 1 & $\alpha^2$ \\ 
\end{tabular}    
\end{pmatrix}
=
\begin{pmatrix}
\begin{tabular}{ll} 
 1 & 0 \\ 
 1 & 1  \\ 
\end{tabular}    
\end{pmatrix}
\begin{pmatrix}
\begin{tabular}{ll} 
 1 & $\alpha$ \\ 
 0 & 1            \\ 
\end{tabular}    
\end{pmatrix}.
$$

\subsubsection{$m=4$}

The finite field $GF(2^4)$ is defined by an element $\alpha$,
which is a root of the primitive polynomial $x^4 + x + 1$. 
Let $c_k$=1, while the binary conjugacy class is 
$\left(\alpha^{1},\alpha^{2},\alpha^{4},\alpha^{8}\right)$ of $GF(2^4)$.

$$V_1  = 
\begin{pmatrix}
\begin{tabular}{llll} 
 1 & $\alpha^1$ & $\alpha^2$ & $\alpha^3$   \\
 1 & $\alpha^2$ & $\alpha^4$ & $\alpha^6$   \\
 1 & $\alpha^4$ & $\alpha^8$ & $\alpha^{12}$\\
 1 & $\alpha^8$ & $\alpha^1$ & $\alpha^9$   \\
\end{tabular}                                
\end{pmatrix}
= 
\begin{pmatrix}
\begin{tabular}{ll|ll} 
 1 & 0 & 0 & 0 \\ 
 0 & 1 & 0 & 0 \\ \hline
 1 & 0 & 1 & 0 \\ 
 0 & 1 & 0 & 1 \\ 
\end{tabular}                                
\end{pmatrix}
\begin{pmatrix}
\begin{tabular}{ll|ll} 
 1 & 0 & $\alpha$ & 0                 \\ 
 0 & 1 & 0             & $\alpha^2$ \\ \hline
 0 & 0 & 1             & 0                  \\ 
 0 & 0 & 0             & 1                  \\ 
\end{tabular}                                
\end{pmatrix}
$$
$$
\times
\begin{pmatrix}
\begin{tabular}{ll|ll} 
 1 & $\alpha^5$       & 0             & 0                         \\ 
 1 & $\alpha^{10}$ & 0             & 0                         \\ \hline
 0 & 0                       & 1              & $\alpha^5$        \\ 
 0 & 0                       & 1              & $\alpha^{10}$  \\ 
\end{tabular}                                
\end{pmatrix}
\begin{pmatrix}
\begin{tabular}{llll} 
 1 & 0 & 0 & 0 \\ 
 0 & 0 & 1 & 0 \\ 
 0 & 1 & 0 & 0 \\ 
 0 & 0 & 0 & 1 \\ 
\end{tabular}                                
\end{pmatrix}
\begin{pmatrix}
\begin{tabular}{llll} 
 1 & 0 & 0 & 0 \\ 
 0 & 1 & 1 & 1 \\ 
 0 & 0 & 1 & 1 \\ 
 0 & 0 & 0 & 1 \\ 
\end{tabular}                                
\end{pmatrix}.
$$

Using $\delta = \alpha^5$, we have 

$$
\Delta = 
\begin{pmatrix}
\begin{tabular}{ll} 
 1 & $\alpha^5$       \\ 
 1 & $\alpha^{10}$ \\ 
\end{tabular}    
\end{pmatrix}
=
\begin{pmatrix}
\begin{tabular}{ll} 
 1 & $\delta$     \\ 
 1 & $\delta^2$ \\ 
\end{tabular}    
\end{pmatrix} = 
\begin{pmatrix}
\begin{tabular}{ll} 
 1 & 0 \\ 
 1 & 1  \\ 
\end{tabular}    
\end{pmatrix}
\begin{pmatrix}
\begin{tabular}{ll} 
 1 & $\delta$ \\ 
 0 & 1            \\ 
\end{tabular}    
\end{pmatrix}.
$$

\section{The novel method for the DFT computation}

Consider several constructions of the novel method on the basis of different algorithms.

The binary conjugacy classes of cardinality $m_i < m$ are considered
for the subfield $GF(2^{m_i}) \subset GF(2^m)$.
If $m_i$ is even then we can apply formula (9) for multipoint evaluation matrix construction.
In other cases we can use the classic methods (for example, \cite{Blahut83}) for the multipoint evaluation 
or normalized cyclic convolution computation.
We further assume that a cardinality of the binary conjugacy classes is $m$.
Let $J_m$ be the indices set of the binary conjugacy classes of cardinality $m$.

According to Theorem 3, we can choose the $k$th binary conjugacy class for which exists the multipoint evaluation matrix $V_k$ (9).

\subsection{The novel method based on the Goertzel--Blahut algorithm}

Consider the construction of the novel method based on the Goertzel--Blahut algorithm 
(formulae (1) and (2)) for even $m$.
From Lemma 3  it follows that for all binary conjugacy classes of cardinality $m$,
we have 
$$V_j = V_k (M_k^T)^{-1} M_j^T,$$
where $j \in J_m$,  
$M_j$ and $M_k$ are the basis transformation matrices.

Using (2), we have 
$$
V = 
\begin{pmatrix}
V_0 &        &            &               \\
       & V_1 &            &               \\
       &        & \ddots &               \\
       &        &            & V_{l-1}  \\
\end{pmatrix}
$$
$$
=
\begin{pmatrix}
\ddots &        &        &            \\
           & V_k &        &             \\
           &        & V_k &             \\
           &        &        & \ddots  \\
\end{pmatrix}
\begin{pmatrix}
\ddots &                                            &                                            &            \\
           & (M_k^T)^{-1} M_{j_1}^T &                                            &            \\
           &                                            & (M_k^T)^{-1} M_{j_2}^T &            \\
           &                                            &                                            & \ddots \\
\end{pmatrix}
= DP,
$$
where $J_m=\{j_1,j_2, \ldots \}$, 
$D$ is the block diagonal matrix of the multipoint evaluation matrices, and 
$P$ is the binary block diagonal matrix of combined preadditions.

In matrix form, the DFT algorithm can be written as  

$$\pi F = V R f = D (P R) f.$$   
where 
$\pi$ is the permutation matrix, 
$D$ is the block diagonal matrix of the multipoint evaluation matrices, 
$P$ is the binary block diagonal matrix of combined preadditions, 
$R$ is the binary matrix.

The algorithm of length $n=2^m-1$ over $GF(2^m)$ takes two steps:

\begin{enumerate}
\item The first step is multiplying the binary matrix $PR$ by the vector $f$
over $GF(2^m)$;
\item The second step is calculation of $l$ $m$-point multipoint evaluations.
\end{enumerate}

\subsection{The novel method based on the cyclotomic algorithm}

Consider the construction of the novel method based on the cyclotomic algorithm 
(formulae (3) and (4)) for even $m$.
From Lemma 2 it follows that for all binary conjugacy classes of cardinality $m$,
we have 
$$L_j = L_k = V_k (M_k^T)^{-1},$$
where $j \in J_m$,  
$M_k$ is the basis transformation matrix.

Using (4), we have 
\begin{equation}
L = 
\begin{pmatrix}
L_0 &        &            &               \\
       & L_1 &            &               \\
       &        & \ddots &               \\
       &        &            & L_{l-1}  \\
\end{pmatrix}
$$
$$
=
\begin{pmatrix}
\ddots &        &        &            \\
           & V_k &        &             \\
           &        & V_k &             \\
           &        &        & \ddots  \\
\end{pmatrix}
\begin{pmatrix}
\ddots &                         &                         &            \\
           & (M_k^T)^{-1} &                         &            \\
           &                         & (M_k^T)^{-1} &            \\
           &                         &                         & \ddots \\
\end{pmatrix}
= DP_c,
\end{equation}
where 
$D$ is the block diagonal matrix of the multipoint evaluation matrices, and 
$P_c$ is the binary block diagonal matrix of combined preadditions.

In matrix form, the DFT algorithm can be written as  

$$F = W (\pi f) = A L (\pi f) = A D P_c (\pi f),$$   
where 
$W = (\alpha^{ij})$, \, $i,j \in [0,n-1]$, is a Vandermonde matrix, 
$A$ is the binary matrix, 
$D$ is the block diagonal matrix of the multipoint evaluation matrices, 
$P_c$ is the binary block diagonal matrix of combined preadditions, 
$\pi$ is the permutation matrix. 

We see that the algorithm contains two multiplications of the binary matrix by the vector.
To reduce the complexity, we consider a modification of this algorithm.

\subsubsection{The novel method based on the inverse cyclotomic algorithm}

Let us write the inverse DFT over the field $GF(2^m)$ in matrix form: 
$$F=W^{-1}(\pi_i f),$$
where $W = (\alpha^{ij})$, \, $i,j \in [0,n-1]$, is a Vandermonde matrix, 
$\pi_i$ is the permutation matrix.

The matrix $L^{-1}$ is a block diagonal matrix composed of basis circulants \cite{Fedorenko04}.

Similarly (11), we have 
$$L^{-1} = D P_i,$$
where 
$D$ is the block diagonal matrix of the multipoint evaluation matrices, and 
$P_i$ is the binary block diagonal matrix of combined preadditions.

Finally, we obtain
$$F = W^{-1}(\pi_i f) = (A L)^{-1}(\pi_i f) = L^{-1} A^{-1}(\pi_i f) = D (P_i A^{-1})(\pi_i f),$$
where 
$D$ is the block diagonal matrix of the multipoint evaluation matrices, 
$P_i$ is the binary block diagonal matrix of combined preadditions, 
$A$ is the binary matrix, 
$\pi_i$ is the permutation matrix. 

The algorithm consists of the multiplication of the binary matrix by the vector and calculation of $l$ $m$-point multipoint evaluations.

\subsubsection{The novel method based on the recurrent algorithm}

The recurrent algorithm \cite{Fedorenko06} is a modification of the cyclotomic algorithm.
This algorithm has a regular structure of the transform matrix $A_r L$.
The multiplicative complexity of the cyclotomic and recurrent algorithms is the same.

Substituting (11) into matrix form of the recurrent algorithm, we have 

$$\pi_r F = A_r L (\pi_r f) = L^T A_r^T (\pi_r f) = L A_r^T (\pi_r f) = D (P_c A_r^T) (\pi_r f),$$
where 
$D$ is the block diagonal matrix of the multipoint evaluation matrices, 
$P_c$ is the binary block diagonal matrix of combined preadditions, 
$A_r$ is a binary matrix with a regular structure, 
$\pi_r$ is the permutation matrix.

The algorithm consists of the multiplication of the binary matrix by the vector and calculation of $l$ $m$-point multipoint evaluations.

\subsection{The complexity of the DFT computation for even extension degree finite field}

It is well-known that the number of the binary conjugacy class of cardinality $i$ and irreducible binary polynomials of degree $i$ 
is the same \cite{MacWilliams}.
The number of irreducible binary polynomials of degree $i$ is 
\cite[Chapter 4, Theorem 15]{MacWilliams}

$$ \Irr(i) = \frac{1}{i} \sum_{d|i} \mu(d) 2^{i/d},$$
where $\mu(d)$ is the M\"obius function.

The number of multiplications of the novel method for the $(2^m-1)$-point DFT computation is
$$\sum_{i|m} \Mult(i) \Irr(i),$$
where $\Mult(m)$ is the number of multiplications of the multipoint evaluation (10).

The complexity of the $n$-point DFT computation for even $m$ of 
the cyclotomic algorithm and novel method is shown in Table \ref{table2}.
The number of multiplications for the cyclotomic algorithm 
for lengths 15, 63, and 255 is cited from the original paper \cite{Fedorenko03},
for lengths 1023 and 4095 it is calculated in the paper \cite[Table 4]{Wu}.

\begin{table}[!t]
\renewcommand{\arraystretch}{1.3}
\caption{The complexity of the $n$-point DFT computation}
\label{table2}
\begin{center}
\begin{tabular}{|r||r|r|}\hline
     & \bfseries cyclotomic algorithm & \bfseries novel method    \\ \hline
$n$  & \bfseries multiplications      & \bfseries multiplications \\ \hline\hline
  15   &      16&     13 \\  
  63   &      97&     88 \\
 255  &     586&    373 \\ \hline
1023 &    2827&   2332 \\
4095 &   10832&   8140\\ \hline
\end{tabular}                                                             
\end{center}
\end{table}

For the multiplication of the binary matrix by the vector,
we can use the modified ``four Russians" algorithm (V. L. Arlazarov, E. A. Dinits, M. A. Kronrod, and I. A. Faradzhev)\cite[Algorithm 6.2]{Aho} 
for multiplication of Boolean matrices, with complexity less than $2 n^2 / \log_2 n$ additions
over elements of $GF(2^m)$. 
On the other hand, we can use the heuristic algorithms
(for example, \cite{Chen, Trifonov07, Trifonov}), whose complexity we could not estimate.
Note that since the matrix of combined preadditions is multiplied by a binary matrix,
it follows that the algorithm complexity does not depend on preadditions.

The asymptotic complexity of the calculation of $l$ $m$-point multipoint evaluations is
$$O(l m \log_2 m) = O\left( \frac{n}{m} m \log_2 m \right) = O(n \log_2\log_2 n)$$
multiplications and additions over elements of $GF(2^m)$.

\section{Examples}

Consider the DFT of length $n = 15$ over the field $GF(2^4)$. 
The finite field $GF(2^4)$ is defined by an element $\alpha$,
which is a root of the primitive polynomial $x^4 + x + 1$. 
Let us take the primitive element $\alpha$ as a transform kernel. 
The binary conjugacy classes of $GF(2^m)$ are: 
$(\alpha^0)$, $(\alpha^1,\alpha^2, \alpha^4, \alpha^8)$, $(\alpha^3, \alpha^6, \alpha^{12}, \alpha^9)$, 
$(\alpha^7, \alpha^{14}, \alpha^{13}, \alpha^{11})$, $(\alpha^5, \alpha^{10})$.

\subsection{The Goertzel--Blahut algorithm}

The first step of the Goertzel--Blahut algorithm is 

%\begin{small}
$$
\begin{tabular}{rcrcr} 
$f(x)$ &=& $(x+1) q_0(x)$             +$r_0(x)$,&$r_0(x)$ = &                         $r_{0,0}$       \\
$f(x)$ &=& $(x^4+x+1) q_1(x)$         +$r_1(x)$,&$r_1(x)$ = &$r_{3,1}x^3 + r_{2,1}x^2 + r_{1,1}x$ + $r_{0,1}$\\
$f(x)$ &=& $(x^4+x^3+x^2+x+1) q_2(x)$ +$r_2(x)$,&$r_2(x)$ = &$r_{3,2}x^3 + r_{2,2}x^2 + r_{1,2}x$ + $r_{0,2}$\\
$f(x)$ &=& $(x^4+x^3+1) q_3(x)$       +$r_3(x)$,&$r_3(x)$ = &$r_{3,3}x^3 + r_{2,3}x^2 + r_{1,3}x$ + $r_{0,3}$\\
$f(x)$ &=& $(x^2+x+1) q_4(x)$         +$r_4(x)$,&$r_4(x)$ = &$r_{1,4}x$  + $r_{0,4}$\\
\end{tabular},
$$
%\end{small}

where
$$
\begin{tabular}{ll} 
$r_{0,0}$ &= $\sum_{i=0}^{14} f_i$ \\ \hline
$r_{0,1}$ &= $f_{0} + f_{4} + f_{7} + f_{8} + f_{10} + f_{12} + f_{13} + f_{14}$ \\
$r_{1,1}$ &= $f_{1} + f_{4} + f_{5} + f_{7} + f_{9 } + f_{10} + f_{11} + f_{12}$ \\
$r_{2,1}$ &= $f_{2} + f_{5} + f_{6} + f_{8} + f_{10} + f_{11} + f_{12} + f_{13}$ \\
$r_{3,1}$ &= $f_{3} + f_{6} + f_{7} + f_{9} + f_{11} + f_{12} + f_{13} + f_{14}$ \\ \hline
$r_{0,2}$ &= $f_{0} + f_{4} + f_{5} + f_{9} + f_{10} + f_{14}$ \\
$r_{1,2}$ &= $f_{1} + f_{4} + f_{6} + f_{9} + f_{11} + f_{14}$ \\
$r_{2,2}$ &= $f_{2} + f_{4} + f_{7} + f_{9} + f_{12} + f_{14}$ \\
$r_{3,2}$ &= $f_{3} + f_{4} + f_{8} + f_{9} + f_{13} + f_{14}$ \\ \hline
$r_{0,3}$ &= $f_{0} + f_{4} + f_{5} + f_{6} + f_{7 }+  f_{9 } + f_{11} + f_{12}$ \\
$r_{1,3}$ &= $f_{1} + f_{5} + f_{6} + f_{7} + f_{8 }+  f_{10} + f_{12} + f_{13}$ \\
$r_{2,3}$ &= $f_{2} + f_{6} + f_{7} + f_{8} + f_{9 }+  f_{11} + f_{13} + f_{14}$ \\
$r_{3,3}$ &= $f_{3} + f_{4} + f_{5} + f_{6} + f_{8 }+  f_{10} + f_{11} + f_{14}$ \\ \hline
$r_{0,4}$ &= $f_{0} + f_{2} + f_{3} + f_{5} + f_{6 }+  f_{8 } + f_{9 } + f_{11} + f_{12} + f_{14}$ \\
$r_{1,4}$ &= $f_{1} + f_{2} + f_{4} + f_{5} + f_{7 }+  f_{8 } + f_{10} + f_{11} + f_{13} + f_{14}$.
\end{tabular}
$$

The second step of the Goertzel--Blahut algorithm is   

$$
\begin{tabular}{lllcr} 
$F_{ 0}$ &= $f(\alpha^0)$    &= $r_{0,0}$           & &                                                                      \\ \hline
$F_{ 1}$ &= $f(\alpha^1)$    &= $r_1(\alpha^1)$    &=& $r_{3,1}\alpha^3    + r_{2,1}\alpha^2    + r_{1,1}\alpha^1    + r_{0,1}$ \\
$F_{ 2}$ &= $f(\alpha^2)$    &= $r_1(\alpha^2)$    &=& $r_{3,1}\alpha^6    + r_{2,1}\alpha^4    + r_{1,1}\alpha^2    + r_{0,1}$ \\
$F_{ 4}$ &= $f(\alpha^4)$    &= $r_1(\alpha^4)$    &=& $r_{3,1}\alpha^{12} + r_{2,1}\alpha^8    + r_{1,1}\alpha^4    + r_{0,1}$ \\
$F_{ 8}$ &= $f(\alpha^8)$    &= $r_1(\alpha^8)$    &=& $r_{3,1}\alpha^9    + r_{2,1}\alpha^1    + r_{1,1}\alpha^8    + r_{0,1}$ \\ \hline
$F_{ 3}$ &= $f(\alpha^{ 3})$ &= $r_2(\alpha^{ 3})$ &=& $r_{3,2}\alpha^9    + r_{2,2}\alpha^6    + r_{1,2}\alpha^3    + r_{0,2}$ \\
$F_{ 6}$ &= $f(\alpha^{ 6})$ &= $r_2(\alpha^{ 6})$ &=& $r_{3,2}\alpha^3    + r_{2,2}\alpha^{12} + r_{1,2}\alpha^6    + r_{0,2}$ \\
$F_{12}$ &= $f(\alpha^{12})$ &= $r_2(\alpha^{12})$ &=& $r_{3,2}\alpha^6    + r_{2,2}\alpha^9    + r_{1,2}\alpha^{12} + r_{0,2}$ \\
$F_{ 9}$ &= $f(\alpha^{ 9})$ &= $r_2(\alpha^{ 9})$ &=& $r_{3,2}\alpha^{12} + r_{2,2}\alpha^3    + r_{1,2}\alpha^9    + r_{0,2}$ \\ \hline
$F_{ 7}$ &= $f(\alpha^{ 7})$ &= $r_3(\alpha^{ 7})$ &=& $r_{3,3}\alpha^6    + r_{2,3}\alpha^{14} + r_{1,3}\alpha^7    + r_{0,3}$ \\
$F_{14}$ &= $f(\alpha^{14})$ &= $r_3(\alpha^{14})$ &=& $r_{3,3}\alpha^{12} + r_{2,3}\alpha^{13} + r_{1,3}\alpha^{14} + r_{0,3}$ \\
$F_{13}$ &= $f(\alpha^{13})$ &= $r_3(\alpha^{13})$ &=& $r_{3,3}\alpha^9    + r_{2,3}\alpha^{11} + r_{1,3}\alpha^{13} + r_{0,3}$ \\
$F_{11}$ &= $f(\alpha^{11})$ &= $r_3(\alpha^{11})$ &=& $r_{3,3}\alpha^{3}  + r_{2,3}\alpha^7    + r_{1,3}\alpha^{11} + r_{0,3}$ \\ \hline
$F_{ 5}$ &= $f(\alpha^{ 5})$ &= $r_4(\alpha^{ 5})$ &=& $                                        r_{1,4}\alpha^{ 5} + r_{0,4}$ \\
$F_{10}$ &= $f(\alpha^{10})$ &= $r_4(\alpha^{10})$ &=& $                                        r_{1,4}\alpha^{10} + r_{0,4}$ \\ 
\end{tabular},
$$

or 

$$F_0 = (1) \, (r_{0,0}) = V_0 \, (r_{0,0}),$$

$$
\begin{pmatrix}
F_1\\                                        
F_2\\
F_4\\
F_8\\
\end{pmatrix}
=
\begin{pmatrix}
\begin{tabular}{llll} 
 1 & $\alpha^1$ & $\alpha^2$ & $\alpha^3$   \\
 1 & $\alpha^2$ & $\alpha^4$ & $\alpha^6$   \\
 1 & $\alpha^4$ & $\alpha^8$ & $\alpha^{12}$\\
 1 & $\alpha^8$ & $\alpha^1$ & $\alpha^9$   \\
\end{tabular}                                
\end{pmatrix}
\begin{pmatrix}
r_{0,1}\\
r_{1,1}\\
r_{2,1}\\
r_{3,1}\\
\end{pmatrix}
= V_1 
\begin{pmatrix}
r_{0,1}\\
r_{1,1}\\
r_{2,1}\\
r_{3,1}\\
\end{pmatrix}, 
$$

$$
\begin{pmatrix}
F_{ 3}\\                                        
F_{ 6}\\
F_{12}\\
F_{ 9}\\
\end{pmatrix}
=
\begin{pmatrix}
\begin{tabular}{llll} 
 1 & $\alpha^3$    & $\alpha^6$    & $\alpha^9$   \\
 1 & $\alpha^6$    & $\alpha^{12}$ & $\alpha^3$   \\
 1 & $\alpha^{12}$ & $\alpha^9$    & $\alpha^6$   \\
 1 & $\alpha^9$    & $\alpha^3$    & $\alpha^{12}$\\
\end{tabular}                                
\end{pmatrix}
\begin{pmatrix}
r_{0,2}\\
r_{1,2}\\
r_{2,2}\\
r_{3,2}\\
\end{pmatrix}
= V_2
\begin{pmatrix}
r_{0,2}\\
r_{1,2}\\
r_{2,2}\\
r_{3,2}\\
\end{pmatrix},
$$

$$
\begin{pmatrix}
F_{ 7}\\                                        
F_{14}\\
F_{13}\\
F_{11}\\
\end{pmatrix}
=
\begin{pmatrix}
\begin{tabular}{llll} 
 1 & $\alpha^7$    & $\alpha^{14}$ & $\alpha^6$    \\
 1 & $\alpha^{14}$ & $\alpha^{13}$ & $\alpha^{12}$ \\            
 1 & $\alpha^{13}$ & $\alpha^{11}$ & $\alpha^9$    \\            
 1 & $\alpha^{11}$ & $\alpha^7$    & $\alpha^3$    \\
\end{tabular}                                
\end{pmatrix}
\begin{pmatrix}
r_{0,3}\\
r_{1,3}\\
r_{2,3}\\
r_{3,3}\\
\end{pmatrix}
= V_3
\begin{pmatrix}
r_{0,3}\\
r_{1,3}\\
r_{2,3}\\
r_{3,3}\\
\end{pmatrix}, 
$$

$$
\begin{pmatrix}
F_{ 5}\\                                        
F_{10}\\
\end{pmatrix}
=
\begin{pmatrix}
\begin{tabular}{ll} 
 1 & $\alpha^5$    \\ 
 1 & $\alpha^{10}$ \\ 
\end{tabular}                                
\end{pmatrix}
\begin{pmatrix}
r_{0,4}\\
r_{1,4}\\
\end{pmatrix}
= V_4
\begin{pmatrix}
r_{0,4}\\
r_{1,4}\\
\end{pmatrix}.
$$

The Goertzel--Blahut algorithm in matrix form is
$$\pi F = V R f$$   
or 
\begin{small}
$$
\begin{pmatrix}
F_{ 0}\\ \hline
F_{ 1}\\
F_{ 2}\\
F_{ 4}\\ 
F_{ 8}\\ \hline
F_{ 3}\\
F_{ 6}\\
F_{12}\\
F_{ 9}\\ \hline
F_{ 7}\\
F_{14}\\
F_{13}\\
F_{11}\\ \hline
F_{ 5}\\
F_{10}\\
\end{pmatrix}=
\begin{pmatrix}
\begin{tabular}{l|llll|llll|llll|ll} 
 $\alpha^0$ &            &               &               &               &&&&&&&&&& \\ \hline
            & $\alpha^0$ & $\alpha^1$    & $\alpha^2$    & $\alpha^3$    &&&&&&&&&& \\
            & $\alpha^0$ & $\alpha^2$    & $\alpha^4$    & $\alpha^6$    &&&&&&&&&& \\
            & $\alpha^0$ & $\alpha^4$    & $\alpha^8$    & $\alpha^{12}$ &&&&&&&&&& \\
            & $\alpha^0$ & $\alpha^8$    & $\alpha^1$    & $\alpha^9$    &&&&&&&&&& \\ \hline
 &&&&&        $\alpha^0$ & $\alpha^3$    & $\alpha^6$    & $\alpha^9$    &&&&&& \\ 
 &&&&&        $\alpha^0$ & $\alpha^6$    & $\alpha^{12}$ & $\alpha^3$    &&&&&& \\       
 &&&&&        $\alpha^0$ & $\alpha^{12}$ & $\alpha^9$    & $\alpha^6$    &&&&&& \\       
 &&&&&        $\alpha^0$ & $\alpha^9$    & $\alpha^3$    & $\alpha^{12}$ &&&&&& \\ \hline
 &&&&&&&&&    $\alpha^0$ & $\alpha^7$    & $\alpha^{14}$ & $\alpha^6$    && \\
 &&&&&&&&&    $\alpha^0$ & $\alpha^{14}$ & $\alpha^{13}$ & $\alpha^{12}$ && \\            
 &&&&&&&&&    $\alpha^0$ & $\alpha^{13}$ & $\alpha^{11}$ & $\alpha^9$    && \\            
 &&&&&&&&&    $\alpha^0$ & $\alpha^{11}$ & $\alpha^7$    & $\alpha^3$    && \\ \hline 
 &&&&&&&&&&&&&$\alpha^0$ & $\alpha^5$    \\
 &&&&&&&&&&&&&$\alpha^0$ & $\alpha^{10}$ \\
\end{tabular}
\end{pmatrix}
$$
\end{small}

$$
\times
\begin{pmatrix}
\begin{tabular}{ccccccccccccccc} 
 1&1&1&1&1&1&1&1&1&1&1&1&1&1&1 \\ \hline 
 1&0&0&0&1&0&0&1&1&0&1&0&1&1&1 \\
 0&1&0&0&1&1&0&1&0&1&1&1&1&0&0 \\
 0&0&1&0&0&1&1&0&1&0&1&1&1&1&0 \\
 0&0&0&1&0&0&1&1&0&1&0&1&1&1&1 \\ \hline 
 1&0&0&0&1&1&0&0&0&1&1&0&0&0&1 \\
 0&1&0&0&1&0&1&0&0&1&0&1&0&0&1 \\
 0&0&1&0&1&0&0&1&0&1&0&0&1&0&1 \\
 0&0&0&1&1&0&0&0&1&1&0&0&0&1&1 \\ \hline 
 1&0&0&0&1&1&1&1&0&1&0&1&1&0&0 \\
 0&1&0&0&0&1&1&1&1&0&1&0&1&1&0 \\
 0&0&1&0&0&0&1&1&1&1&0&1&0&1&1 \\
 0&0&0&1&1&1&1&0&1&0&1&1&0&0&1 \\ \hline 
 1&0&1&1&0&1&1&0&1&1&0&1&1&0&1 \\
 0&1&1&0&1&1&0&1&1&0&1&1&0&1&1 \\
\end{tabular}
\end{pmatrix}
\begin{pmatrix}
f_{ 0}\\ 
f_{ 1}\\
f_{ 2}\\
f_{ 3}\\ 
f_{ 4}\\
f_{ 5}\\
f_{ 6}\\ 
f_{ 7}\\
f_{ 8}\\
f_{ 9}\\
f_{10}\\
f_{11}\\
f_{12}\\
f_{13}\\
f_{14}\\
\end{pmatrix}.
$$

\subsection{The novel method based on the Goertzel--Blahut algorithm}

In the field $GF(2^4)$, the normal basis is 
$\left(\gamma^1,\gamma^2,\gamma^4,\gamma^8 \right)$ = 
$\left(\alpha^{6},\alpha^{12},\alpha^{9},\alpha^{3} \right)$;
in its subfields, the normal bases are  
$\left(\alpha^{5},\alpha^{10} \right)$ for $GF(2^2) \subset GF(2^4)$, 
and $\left(\alpha^{0} \right)$ for $GF(2) \subset GF(2^4)$. 

Let us denote the basis circulants by 

$$
L_0 = 
\begin{pmatrix}
 1 \\
\end{pmatrix}, \quad
L_1 = L_2 = L_3 = 
\begin{pmatrix}
 \gamma^{ 1} & \gamma^{ 2} & \gamma^{ 4} & \gamma^{ 8} \\
 \gamma^{ 2} & \gamma^{ 4} & \gamma^{ 8} & \gamma^{ 1} \\
 \gamma^{ 4} & \gamma^{ 8} & \gamma^{ 1} & \gamma^{ 2} \\
 \gamma^{ 8} & \gamma^{ 1} & \gamma^{ 2} & \gamma^{ 4} \\
\end{pmatrix}, \quad
L_4 = 
\begin{pmatrix}
\alpha^{5}   & \alpha^{10} \\
\alpha^{10} & \alpha^{5}   \\
\end{pmatrix}.
$$

Using $V_k = L_k M_k^T$, we have the basis transformation matrices $M_k$, $k \in [0,4]$:

$$
(M_0)^T = 
\begin{pmatrix}
\begin{tabular}{c} 
 1 \\
\end{tabular}
\end{pmatrix},
\qquad
(M_1)^T = 
\begin{pmatrix}
\begin{tabular}{cccc} 
 1&0&1&0 \\
 1&0&0&0 \\
 1&1&0&0 \\
 1&1&1&1 \\
\end{tabular}
\end{pmatrix},
\qquad
(M_2)^T = 
\begin{pmatrix}
\begin{tabular}{cccc} 
 1&0&1&0 \\
 1&0&0&0 \\
 1&0&0&1 \\
 1&1&0&0 \\
\end{tabular}
\end{pmatrix},
$$

$$
(M_3)^T = 
\begin{pmatrix}
\begin{tabular}{cccc} 
 1&1&1&1 \\
 1&1&1&0 \\
 1&0&1&0 \\
 1&1&0&0 \\
\end{tabular}
\end{pmatrix},
\qquad
(M_4)^T = 
\begin{pmatrix}
\begin{tabular}{cc} 
 1&1 \\
 1&0 \\
\end{tabular}
\end{pmatrix}.
$$

Let us remember that the multipoint evaluation matrix 
for the binary conjugacy class
$\left(\alpha^{1},\alpha^{2},\alpha^{4},\alpha^{8}\right)$ is:

$$V_1  = 
\begin{pmatrix}
\begin{tabular}{llll} 
 1 & $\alpha^1$ & $\alpha^2$ & $\alpha^3$   \\
 1 & $\alpha^2$ & $\alpha^4$ & $\alpha^6$   \\
 1 & $\alpha^4$ & $\alpha^8$ & $\alpha^{12}$\\
 1 & $\alpha^8$ & $\alpha^1$ & $\alpha^9$   \\
\end{tabular}                                
\end{pmatrix}
= 
\begin{pmatrix}
\begin{tabular}{ll|ll} 
 1 & 0 & 0 & 0 \\ 
 0 & 1 & 0 & 0 \\ \hline
 1 & 0 & 1 & 0 \\ 
 0 & 1 & 0 & 1 \\ 
\end{tabular}                                
\end{pmatrix}
\begin{pmatrix}
\begin{tabular}{ll|ll} 
 1 & 0 & $\alpha$ & 0                 \\ 
 0 & 1 & 0             & $\alpha^2$ \\ \hline
 0 & 0 & 1             & 0                  \\ 
 0 & 0 & 0             & 1                  \\ 
\end{tabular}                                
\end{pmatrix}
$$
$$
\times
\begin{pmatrix}
\begin{tabular}{ll|ll} 
 1 & $\alpha^5$       & 0             & 0                         \\ 
 1 & $\alpha^{10}$ & 0             & 0                         \\ \hline
 0 & 0                       & 1              & $\alpha^5$        \\ 
 0 & 0                       & 1              & $\alpha^{10}$  \\ 
\end{tabular}                                
\end{pmatrix}
\begin{pmatrix}
\begin{tabular}{llll} 
 1 & 0 & 0 & 0 \\ 
 0 & 0 & 1 & 0 \\ 
 0 & 1 & 0 & 0 \\ 
 0 & 0 & 0 & 1 \\ 
\end{tabular}                                
\end{pmatrix}
\begin{pmatrix}
\begin{tabular}{llll} 
 1 & 0 & 0 & 0 \\ 
 0 & 1 & 1 & 1 \\ 
 0 & 0 & 1 & 1 \\ 
 0 & 0 & 0 & 1 \\ 
\end{tabular}                                
\end{pmatrix} 
$$
$$
= 
\begin{pmatrix}
\begin{tabular}{llll} 
$\alpha^0$ & $\alpha^{ 5}$ & $\alpha^1$ & $\alpha^{ 6}$ \\
$\alpha^0$ & $\alpha^{10}$ & $\alpha^2$ & $\alpha^{12}$ \\
$\alpha^0$ & $\alpha^{ 5}$ & $\alpha^4$ & $\alpha^{ 9}$ \\
$\alpha^0$ & $\alpha^{10}$ & $\alpha^8$ & $\alpha^{ 3}$ \\ 
\end{tabular}                                
\end{pmatrix}
\begin{pmatrix}
\begin{tabular}{cccc} 
 1&0&0&0 \\
 0&0&1&1 \\
 0&1&1&1 \\
 0&0&0&1 \\
\end{tabular}                                
\end{pmatrix}
=S P_1,
$$
where $P_1$ is the matrix of preadditions.

From Lemma 3 it follows that 
$$
\begin{matrix}
\begin{tabular}{cccc} 
$V_0$ =&                           &                             & 1       \\ 
$V_1$ =&                           &                             & $S P_1$ \\ 
$V_2$ =& $V_1 (M_1^T)^{-1} M_2^T$ =& $S P_1 (M_1^T)^{-1} M_2^T$ =& $S P_2$ \\ 
$V_3$ =& $V_1 (M_1^T)^{-1} M_3^T$ =& $S P_1 (M_1^T)^{-1} M_3^T$ =& $S P_3$ \\
\end{tabular}                                
\end{matrix},
$$
where $P_k, \, k \in [1,3]$, is the matrix of preadditions, and 
$M_k, \, k \in [1,3]$, is the basis transformation matrix.

The multiplication of the $2 \times 2$ Moore--Vandermonde matrix $V_4$ 
by the vector is very simple:
$$
\begin{pmatrix}
F_{ 5}\\                                        
F_{10}\\
\end{pmatrix}
=
\begin{pmatrix}
\begin{tabular}{ll} 
 1 & $\alpha^5$    \\ 
 1 & $\alpha^{10}$ \\ 
\end{tabular}                                
\end{pmatrix}
\begin{pmatrix}
r_{0,4}\\
r_{1,4}\\
\end{pmatrix}
$$
$$
= V_4 
\begin{pmatrix}
r_{0,4}\\
r_{1,4}\\
\end{pmatrix}
=
\begin{pmatrix}
\begin{tabular}{ll} 
 1 & 0 \\ 
 1 & 1  \\ 
\end{tabular}    
\end{pmatrix}
\begin{pmatrix}
\begin{tabular}{ll} 
 1 & $\alpha^5$ \\ 
 0 & 1                 \\ 
\end{tabular}    
\end{pmatrix}
\begin{pmatrix}
r_{0,4}\\
r_{1,4}\\
\end{pmatrix}.
$$

Using (2), we obtain
$$
V = 
\begin{pmatrix}
 V_0 &     &     &     &     \\
     & V_1 &     &     &     \\
     &     & V_2 &     &     \\
     &     &     & V_3 &     \\
     &     &     &     & V_4 \\
\end{pmatrix}
$$
$$
=
\begin{pmatrix}
 1 &   &   &   &     \\
   & S &   &   &     \\
   &   & S &   &     \\
   &   &   & S &     \\
   &   &   &   & V_4 \\
\end{pmatrix}
\begin{pmatrix}
 1 &     &     &     &     \\
   & P_1 &     &     &     \\
   &     & P_2 &     &     \\
   &     &     & P_3 &     \\
   &     &     &     & I_2 \\
\end{pmatrix}
= DP,
$$
where $I_2$ is the $2 \times 2$ identity matrix, 
$D$ is the block diagonal matrix of the multipoint evaluation matrices without preadditions, and 
$P$ is the binary block diagonal matrix of combined preadditions.

In matrix form, the DFT algorithm can be written as  
$$\pi F = V R f = D (P R) f$$   
or
\begin{small}
$$
\begin{pmatrix}
F_{ 0}\\ \hline
F_{ 1}\\
F_{ 2}\\
F_{ 4}\\ 
F_{ 8}\\ \hline
F_{ 3}\\
F_{ 6}\\
F_{12}\\
F_{ 9}\\ \hline
F_{ 7}\\
F_{14}\\                          
F_{13}\\                          
F_{11}\\ \hline                   
F_{ 5}\\                          
F_{10}\\
\end{pmatrix}=
\begin{pmatrix}
\begin{tabular}{l|llll|llll|llll|ll} 
 $\alpha^0$ &            &               &            &               &&&&&&&&&& \\ \hline
            & $\alpha^0$ & $\alpha^{ 5}$ & $\alpha^1$ & $\alpha^{ 6}$ &&&&&&&&&& \\
            & $\alpha^0$ & $\alpha^{10}$ & $\alpha^2$ & $\alpha^{12}$ &&&&&&&&&& \\
            & $\alpha^0$ & $\alpha^{ 5}$ & $\alpha^4$ & $\alpha^{ 9}$ &&&&&&&&&& \\
            & $\alpha^0$ & $\alpha^{10}$ & $\alpha^8$ & $\alpha^{ 3}$ &&&&&&&&&& \\ \hline
 &&&&&        $\alpha^0$ & $\alpha^{ 5}$ & $\alpha^1$ & $\alpha^{ 6}$ &&&&&& \\ 
 &&&&&        $\alpha^0$ & $\alpha^{10}$ & $\alpha^2$ & $\alpha^{12}$ &&&&&& \\       
 &&&&&        $\alpha^0$ & $\alpha^{ 5}$ & $\alpha^4$ & $\alpha^{ 9}$ &&&&&& \\       
 &&&&&        $\alpha^0$ & $\alpha^{10}$ & $\alpha^8$ & $\alpha^{ 3}$ &&&&&& \\ \hline
 &&&&&&&&&    $\alpha^0$ & $\alpha^{ 5}$ & $\alpha^1$ & $\alpha^{ 6}$ && \\
 &&&&&&&&&    $\alpha^0$ & $\alpha^{10}$ & $\alpha^2$ & $\alpha^{12}$ && \\            
 &&&&&&&&&    $\alpha^0$ & $\alpha^{ 5}$ & $\alpha^4$ & $\alpha^{ 9}$ && \\            
 &&&&&&&&&    $\alpha^0$ & $\alpha^{10}$ & $\alpha^8$ & $\alpha^{ 3}$ && \\ \hline 
 &&&&&&&&&&&&&$\alpha^0$ & $\alpha^5$    \\
 &&&&&&&&&&&&&$\alpha^0$ & $\alpha^{10}$ \\
\end{tabular}
\end{pmatrix}
$$
\end{small}

$$
\times
\begin{pmatrix}
\begin{tabular}{ccccccccccccccc} 
 1&1&1&1&1&1&1&1&1&1&1&1&1&1&1\\
 1&0&0&0&1&0&0&1&1&0&1&0&1&1&1\\
 0&0&1&1&0&1&0&1&1&1&1&0&0&0&1\\
 0&1&1&1&1&0&0&0&1&0&0&1&1&0&1\\
 0&0&0&1&0&0&1&1&0&1&0&1&1&1&1\\
 1&0&0&0&1&1&0&0&0&1&1&0&0&0&1\\
 0&1&0&1&0&0&1&0&1&0&0&1&0&1&0\\
 0&1&0&0&1&0&1&0&0&1&0&1&0&0&1\\
 0&1&1&1&1&0&1&1&1&1&0&1&1&1&1\\
 1&1&1&0&1&0&1&1&0&0&1&0&0&0&1\\
 0&1&1&0&0&1&0&0&0&1&1&1&1&0&1\\
 0&0&1&0&0&0&1&1&1&1&0&1&0&1&1\\
 0&1&1&1&1&0&1&0&1&1&0&0&1&0&0\\
 1&0&1&1&0&1&1&0&1&1&0&1&1&0&1\\
 0&1&1&0&1&1&0&1&1&0&1&1&0&1&1\\
\end{tabular}
\end{pmatrix}
\begin{pmatrix}
f_{ 0}\\ 
f_{ 1}\\
f_{ 2}\\
f_{ 3}\\ 
f_{ 4}\\
f_{ 5}\\
f_{ 6}\\ 
f_{ 7}\\
f_{ 8}\\
f_{ 9}\\
f_{10}\\
f_{11}\\
f_{12}\\
f_{13}\\
f_{14}\\
\end{pmatrix}.
$$

\subsection{The complexity of the 15-point DFT computation}

The novel method of the 15-point DFT computation based on the Goertzel--Blahut algorithm consists of three parts:

\begin{enumerate}
\item triple multiplication by the matrix $S$
(by the Moore--Vandermonde multipoint evaluation matrix $V_1 = S P_1$ without preadditions):
$3 \times 4 = 12$ multiplications and $3 \times 8 = 24$ additions;

\item multiplication by the matrix $V_4$:
1 multiplication and 2 additions;

\item multiplication of the binary matrix $PR$ by the vector $f$
(using a heuristic algorithm \cite{Trifonov07}):
44 additions.

\end{enumerate}

\noindent
The complexity of this method is 13 multiplications and 70 additions.

The complexity of some methods of the 15-point DFT computation
is shown in Table \ref{table3}.

\begin{table}[!t]
\renewcommand{\arraystretch}{1.3}
\caption{The complexity of the 15-point DFT computation}
\label{table3}
\begin{center}
\centering
\begin{tabular}{|l||c|c|}\hline
\bfseries method & \bfseries multipli- & \bfseries addi- \\
                 & \bfseries cations   & \bfseries tions \\
\hline\hline
the cyclotomic algorithm \cite{Fedorenko03}      & 16 & 77 \\ \hline
the cyclotomic algorithm using                             &      &    \\
an improved heuristic algorithm \cite{Trifonov}   & 16 & 76 \\ \hline
the recurrent algorithm \cite{Fedorenko06}         & 16 & 91 \\ \hline
the cyclotomic algorithm with common                 &     &    \\
subexpression elimination algorithm \cite{Chen} & 16 & 74 \\ \hline
the novel method based on                                   &      &  \\ 
the Goertzel--Blahut algorithm                              & 13 & 70 \\ \hline
the novel method based on                                   &      &  \\ 
the recurrent algorithm                                     & 13 & 68 \\ \hline
\end{tabular}
\end{center}
\end{table}

\section{Conclusion}

A novel method for computation of the DFT 
over a finite field with reduced multiplicative complexity is described.
The method is applicable for any DFT length
but reducing the multiplicative complexity is achieved not for all lengths.
For the DFT computation of length $n=2^m-1$ in the field $GF(2^m)$ for even $m$ 
the novel method is the best known method 
(if the number of multiplications is to be minimized), 
and the exact formula for the number of multiplications is analytically obtained.

\section*{Acknowledgment}
The author would like to thank the Alexander von Humboldt Foundation, Germany, 
for the many years' support of his research, 
Prof. Anja Klein for her hospitality at the Technische Universit\"at Darmstadt, Darmstadt, Germany,
Valentin Afanassiev, and Peter Trifonov for helpful discussions.

\end{document}